\documentclass[journal=cmatex,manuscript=article]{achemso}

\usepackage{chemformula} 
\usepackage[T1]{fontenc} 
\usepackage{pifont}
\usepackage{multirow}
\setkeys{acs}{maxauthors = 0}

\author{Monica {Ciomaga Hatnean}}
\affiliation{Department of Physics, University of Warwick, Coventry CV4 7AL, UK.}
\email{M.Ciomaga-Hatnean@warwick.ac.uk}
\author{Oleg A. Petrenko}
\affiliation{Department of Physics, University of Warwick, Coventry CV4 7AL, UK.}
\author{Martin R. Lees}
\affiliation{Department of Physics, University of Warwick, Coventry CV4 7AL, UK.}
\author{Tom E. Orton}
\affiliation{Department of Physics, University of Warwick, Coventry CV4 7AL, UK.}
\author{Geetha Balakrishnan}
\affiliation{Department of Physics, University of Warwick, Coventry CV4 7AL, UK.}
\title[]
  {Optical floating zone crystal growth of rare-earth disilicates, $R\mathbf{_{2}}$Si$\mathbf{_{2}}$O$\mathbf{_{7}}$ ($R$~=~Er, Ho, and Tm)}
\abbreviations{TBC (Thermal barrier coating), EBC (Environmental barrier coating), CVT (Chemical vapour transport), FZ (Floating zone), QDM (Quantum dimer magnet), BEC (Bose-Einstein condensate), GOF (Goodness of fit).}
\begin{document}

%
%
%
%
%
%

\begin{abstract}
The wealth of structural phases seen in the rare-earth disilicate compounds promises an equally rich range of interesting magnetic properties. We report on the crystal growth by the optical floating zone method of members of the rare-earth disilicate family, $R_{2}$Si$_{2}$O$_{7}$ (with \textit{R}~=~Er, Ho, and Tm). Through a systematic study, we have optimised the growth conditions for Er$_{2}$Si$_{2}$O$_{7}$. We have grown, for the first time using the floating zone method, crystal boules of Ho$_{2}$Si$_{2}$O$_{7}$ and Tm$_{2}$Si$_{2}$O$_{7}$ compounds. We show that the difficulties encountered in the synthesis of polycrystalline and single crystal samples are due to the similar thermal stability ranges of different rare-earth silicate compounds in the temperature-composition phase diagrams of the $R$-Si-O systems. The addition of a small amount of SiO$_{2}$ excess allowed the amount of impurity phases present in the powder samples to be minimised. The phase composition analysis of the powder X-ray diffraction data collected on the as-grown boules revealed that they were of single phase, except in the case of thulium disilicate, which comprised of two phases. All growths resulted in multi-grain boules, from which sizeable single crystals could be isolated. The optimum conditions used for the synthesis and crystal growth of polycrystalline and single crystal $R_{2}$Si$_{2}$O$_{7}$ materials are reported. Specific heat measurements of erbium and thulium disilicate compounds confirm an antiferromagnetic phase transition below $T_{\mathrm{N}}$~=~1.8~K for D-type Er$_{2}$Si$_{2}$O$_{7}$ and a Schottky anomaly centred around 3.5~K in C-type Tm$_{2}$Si$_{2}$O$_{7}$, suggesting the onset of short-range magnetic correlations. Magnetic susceptibility data of E-type Ho$_{2}$Si$_{2}$O$_{7}$ reveals an antiferromagnetic ordering of the Ho spins below $T_\mathrm{{N}}$~=~2.3~K.
\end{abstract}
\section{Introduction}
Rare-earth disilicates, $R_{2}$Si$_{2}$O$_{7}$, where \textit{R} is a rare-earth element, were studied in the past due to the polymorphism exhibited by these compounds and their physical properties. At ambient pressure, $R_{2}$Si$_{2}$O$_{7}$ compounds can crystallise in seven different structural types~\cite{1968_Ito,1970_Felsche,1973_Felsche}. The seven polymorphs, conventionally referred to as A, B (or $\alpha$), C (or $\beta$), D (or $\gamma$), E (or $\delta$), F, G, have, respectively, tetragonal ($P$4$_{1}$), triclinic ($P\bar{1}$), monoclinic ($C$2/$m$), monoclinic ($P$2$_{1}$/$a$), orthorhombic ($Pnam$ or $Pna2_{1}$), triclinic ($P\bar{1}$), and monoclinic ($P$2$_{1}$/$c$) crystal structure~\cite{1970_Felsche}. At room temperature and ambient pressure, $R_{2}$Si$_{2}$O$_{7}$ compounds can be stabilised in one of three of these structures, depending on the ionic radius of the rare-earth element, $R$. Rare-earth disilicate compounds containing large rare-earth elements (with $R~=~$La~$\rightarrow$~Sm) crystallise in the tetragonal A-type structure. Compounds $R_{2}$Si$_{2}$O$_{7}$ that incorporate smaller rare-earth ions (where $R~=~$Eu~$\rightarrow$~Tm) adopt a triclinic B-type structure. The smallest rare-earth elements, Lu and Yb, form disilicate compounds crystallising in the monoclinic C-type structure. At high temperature (950~$\leq~T~\leq$~1500~$^{\circ}$C), rare-earth disilicate compounds $R_{2}$Si$_{2}$O$_{7}$ (where $R~=~$La~$\rightarrow$~Tm) undergo one or more structural phase transitions. The number of phase transitions they undergo and their transition temperatures depend on the nature of the rare-earth ion~\cite{1970_Felsche}. A detailed diagram describing the thermal area of stability of the polymorphs, as well as the characteristics of each crystallographic structure, for each rare-earth disilicate compound, can be found in the early work of Felsche~\cite{1970_Felsche}. Two new polymorphs, K and L, with monoclinic ($P$2$_{1}$/$n$) and triclinic ($P\bar{1}$) crystal symmetries, were later synthesised under pressure~\cite{2002_Liu}.

$R_{2}$Si$_{2}$O$_{7}$ compounds are widely investigated for their luminescent and optical properties, with potential applications as crystal scintillators and in the detection of $\gamma$ and X-rays~\cite{1980_Bretheau-Raynal,2000_Pauwels,2010_Feng,2012_He,2013_Fernandez-Carrion}. Recently, rare-earth disilicates were identified as promising candidates for thermal barrier coating/environmental barrier coating (TBC/EBC) systems, due to their low thermal conductivities~\cite{2005_Lee,2008_Sun,2009_Sun,2013_Zhou,2016_Tian,2018_Luo}. 

Other properties, such as the magnetic behaviour of $R_{2}$Si$_{2}$O$_{7}$ compounds, have not been investigated in great detail, partly due to the complexity of the structural phase diagram. The successful growth of ytterbium and erbium disilicate crystals using optical FZ method~\cite{2019_Nair} has resulted in a resurgence of interest in the magnetic ground states of these compounds~\cite{2019_Hester,2020_Flynn}. Yb$_{2}$Si$_{2}$O$_{7}$ is the first Yb$^{3+}$ based material that exhibits a quantum dimer magnet (QDM) state, consisting of nearest neighbour spin dimers, with magnetic field-induced order reminiscent of a Bose-Einstein condensate (BEC) phase. Nevertheless, ytterbium disilicate is one of the very few $R_{2}$Si$_{2}$O$_{7}$ compounds that crystallises in only one structural type~\cite{1970_Felsche}, whereas for the other members of the rare-earth disilicate compounds stabilising in two or more crystallographic structures, extensive studies of the magnetic properties are scarce in the literature. 

The growth of $R_{2}$Si$_{2}$O$_{7}$ crystals opens up a route to further investigation of these materials, with the potential to unearth some very interesting magnetic properties, similar to what has been observed in Yb$_{2}$Si$_{2}$O$_{7}$. There have been many attempts to grow crystals of the rare-earth disilicate compounds. Typical routes used to prepare $R_{2}$Si$_{2}$O$_{7}$ crystals include the Verneuil~\cite{1970_Smolin}, flux~\cite{1974_Wanklyn,1978_Wanklyn,1979_Maqsood,1997_NorlundChristensen,2000_Maqsood,2014_Kahlenberg}, Czochralski~\cite{2012_He}, chemical vapour transport (CVT)~\cite{1998_Chi}, floating zone (FZ)~\cite{2000_Pauwels,2019_Nair} and micro-pulling-down ($\mu$-PD)~\cite{2016_Horiai} techniques. The FZ method of crystal growth has been widely employed in the past for the growth of oxides, and not only is it the ideal technique to produce large, high quality single crystals, but it is also one of the most appropriate methods due to its advantages compared to the conventional techniques of crystal growth~\cite{1998_Balakrishnan,2008_Koohpayeh,2015_Dabkowska}. The main benefits of employing the standard FZ technique for the growth of crystals are: (a) the high purity of the crystals grown thanks to the absence of a container (crucible) and solvent (flux), (b) the relatively large size of the crystal boules obtained compared with other techniques, (c) the good crystalline quality and uniformity of the physical properties throughout the crystals, (d) this method can be used for growing refractory materials.

This has motivated us to embark upon the study of the highly polymorphic rare-earth disilicate compounds. In the present work, we investigate three rare-earth disilicate compounds, $R_{2}$Si$_{2}$O$_{7}$, where \textit{R}~=~Er, Ho, and Tm, with focus on the crystal growth of these materials using the optical FZ method. Tm$_{2}$Si$_{2}$O$_{7}$ is dimorphic with temperature, whereas Er$_{2}$Si$_{2}$O$_{7}$ and Ho$_{2}$Si$_{2}$O$_{7}$ display a high degree of polymorphism~\cite{1970_Felsche,1987_Maqsood,1997_Maqsood,2000_Maqsood,2009_Maqsood,2013_Fernandez-Carrion,2014_Kahlenberg}. A list of the various polymorphs of the disilicate compounds under investigation in this study along with the temperature ranges of their formation/stability, and their respective crystal structures and melting points is given in Table~\ref{Tab:R2Si2O7_temp_struct}.

Rare-earth disilicate compounds have previously been grown in crystal form using the flux method~\cite{1974_Wanklyn,1978_Wanklyn,1979_Maqsood,1997_NorlundChristensen,2000_Maqsood,2014_Kahlenberg}, however, the temperature regions where only one crystallographic phase forms are narrow and small differences (of 50 to $100~^{\circ}$C) in the soak temperature could result in a yield of crystals of two different polymorphs or in a different structural phase to the one desired~\cite{1987_Maqsood,2009_Maqsood}. Previous studies~\cite{2016_CiomagaHatnean,2017_Sibille} of other materials that undergo thermally-induced structural phase transitions have proven that the stabilisation of a specific structural type is feasible in FZ grown crystals. The crystal growth of the high temperature stable polymorphs of rare-earth disilicate compounds is therefore made possible by the FZ method, especially for those structural types where the melting points (1700~$\leq~T_{m}~\leq$~1800~$^{\circ}$C) are higher than the temperatures of the structural phase transitions (950~$\leq~T~\leq$~1500~$^{\circ}$C)~\cite{1970_Felsche}.

\begin{table}[H]
\caption{Melting points and regions of thermal stability~\cite{1970_Felsche} of each polymorph of the $R_{2}$Si$_{2}$O$_{7}$ (with \textit{R}~=~Er, Ho, and Tm) compounds.}
\small 
\centering
\begin{tabular}{ccccc}
\cline{1-5}
\multirow{2}{*}{\textbf{$\mathbf{R_{2}}$Si$\mathbf{_{2}}$O$\mathbf{_{7}}$}}&\multirow{2}{*}{\textbf{Polymorph}}&\textbf{Space}&\textbf{Thermal stability}&\textbf{Melting}\\
& & \textbf{group} &\textbf{range}&\textbf{point ($T_{m}$)}\\
\cline{1-5}
\multirow{3}{*}{Er$_{2}$Si$_{2}$O$_{7}$}&B-type&$P\bar{1}$&$T~<$~1025~$^{\circ}$C&\multirow{3}{*}{$\sim~1750~^{\circ}$C}\\
	&C-type&$C$2/$m$&1025~$<~T~\leq$~1400~$^{\circ}$C&\\
	&D-type&$P$2$_{1}$/$b$&$T~\geq$~1350~$^{\circ}$C&\\
\cline{1-5}
\multirow{4}{*}{Ho$_{2}$Si$_{2}$O$_{7}$}&B-type&$P\bar{1}$&$T~\leq$~1200~$^{\circ}$C&\multirow{4}{*}{$\sim~1750~^{\circ}$C}\\
	&C-type&$C$2/$m$&1200~$\leq~T~<$~1275~$^{\circ}$C&\\
	&D-type&$P$2$_{1}$/$b$&1275~$<~T~\leq$~1500~$^{\circ}$C&\\
	&E-type&$Pna2_{1}$&$T~\geq$~1500~$^{\circ}$C&\\
	\cline{1-5}
\multirow{2}{*}{Tm$_{2}$Si$_{2}$O$_{7}$}&B-type&$P\bar{1}$&$T~<$~950~$^{\circ}$C&\multirow{2}{*}{$\sim~1700~^{\circ}$C}\\
	&C-type&$C$2/$m$&$T~>$~950~$^{\circ}$C&\\
\cline{1-5}
\end{tabular}
\label{Tab:R2Si2O7_temp_struct}
\end{table}

\begin{figure}[h!]
\centering
\includegraphics[width=0.75\columnwidth]{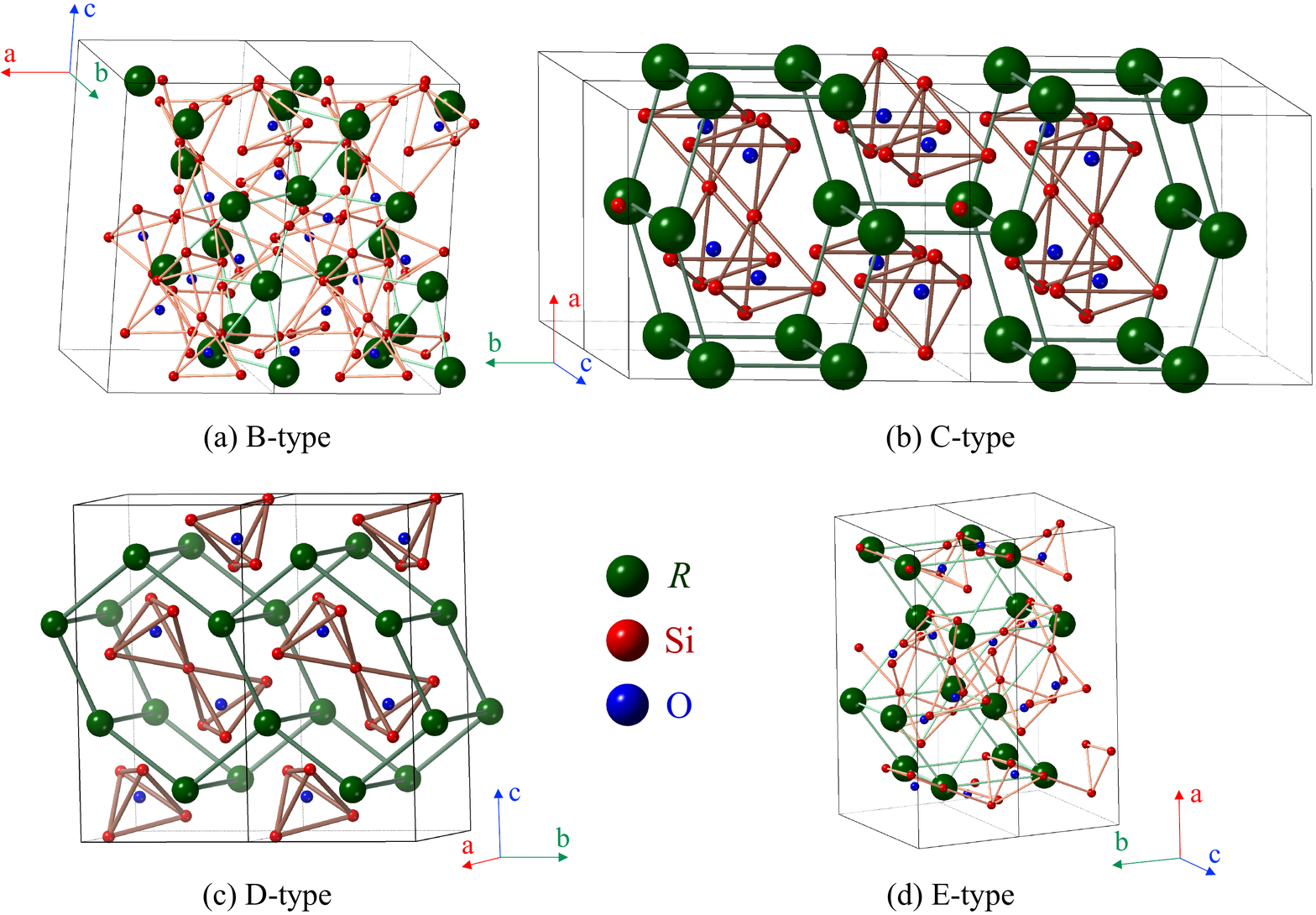}
\caption{Possible crystallographic structures of the $R_{2}$Si$_{2}$O$_{7}$ polymorphs for \textit{R}~=~Er, Ho, and Tm: (a) triclinic ($P\bar{1}$) B-type, (b) monoclinic ($C$2/$m$) C-type, (c) monoclinic ($P$2$_{1}$/$b$) D-type, (d) orthorhombic ($Pna2_{1}$) E-type. The unit cells are shown in black. The bonds between the magnetic $R^{3+}$ ions emphasise the formation of the distorted honeycomb layers in the C-type and D-type $R_{2}$Si$_{2}$O$_{7}$ structures, stacked along the $c$ axis and $a$ axis, respectively. In the E-type structure, the $R^{3+}$ ions form a distorted triangular network.}
\label{Fig:R2Si2O7_structure}
\end{figure}

Erbium disilicate exists in three polymorphs (see Table~\ref{Tab:R2Si2O7_temp_struct}), a triclinic ($P\bar{1}$) low temperature phase (referred to as B-type), a monoclinic ($C$2/$m$) structure (C-type), and a high temperature monoclinic
($P$2$_{1}$/$b$) arrangement (D-type)~\cite{1970_Felsche,1987_Maqsood,1997_Maqsood}. C-type Er$_{2}$Si$_{2}$O$_{7}$ is reported to order antiferromagnetically below $T_\mathrm{{N}}$~=~2.50(5)~K~\cite{1979_Maqsood,2000_Maqsood}. The structural modification stable at high temperature, D-type, is one of the most studied polymorphs of Er$_{2}$Si$_{2}$O$_{7}$, from a magnetic point of view~\cite{1981_Maqsood,1986_Leask}. Furthermore, C~$\rightarrow$~D-type structural phase transition occurs at a lower temperature ($T_{C~\rightarrow~D}~\approx~1425~^{\circ}$C) than the melting point ($T_{m}~\approx~1750~^{\circ}$C) (as shown by Felsche~\cite{1970_Felsche}, all $R_{2}$Si$_{2}$O$_{7}$ compounds do not melt below $1700~^{\circ}$C). A crystal of Er$_{2}$Si$_{2}$O$_{7}$ grown from the melt, by the FZ method will therefore belong to the structural type D~\cite{2019_Nair}. Due to the arrangement of the magnetic ions in the crystallographic structure, forming a distorted honeycomb-like lattice (see Fig.~\ref{Fig:R2Si2O7_structure}(c)), one expects to observe an interesting magnetic behaviour in D-type Er$_{2}$Si$_{2}$O$_{7}$, similar to the case of C-type Yb$_{2}$Si$_{2}$O$_{7}$~\cite{2019_Nair,2019_Hester,2020_Flynn}. Previous reports have shown that D-type Er$_{2}$Si$_{2}$O$_{7}$ orders antiferromagnetically below $T_\mathrm{{N}}$~=~1.9(1)~K, as well as exhibiting a highly anisotropic magnetic behaviour~\cite{1981_Maqsood,1986_Leask}. In addition, these first measurements of the magnetic properties suggest that the application of a strong magnetic field induces a "spin flip"; one or two "flips" are observed depending upon the direction of the applied magnetic field. Detailed investigations are required to determine the magnetic ground state of D-type Er$_{2}$Si$_{2}$O$_{7}$, and to confirm the four-sublattice antiferromagnetic spin model used for calculating the exchange interaction constants in the initial studies.

Holmium disilicate exists in four polymorphs (see Table~\ref{Tab:R2Si2O7_temp_struct}), two low temperature phases (triclinic B-type arrangement, and monoclinic C-type structure), and two high temperature arrangements (monoclinic D-type phase, and an orthorhombic ($Pna2_{1}$) E-type structure)~\cite{1970_Felsche,2009_Maqsood}. The temperatures of the structural phase transitions are close to one another in the low temperature region ($T_{B~\rightarrow~C}~\approx~1200~^{\circ}$C, and $T_{C~\rightarrow~D}~\approx~1275~^{\circ}$C)~\cite{1970_Felsche}, and one can thus expect the synthesis of the low temperature structural types of Ho$_{2}$Si$_{2}$O$_{7}$ to be challenging. A study by Maqsood~\cite{2009_Maqsood} showed that the synthesis attempts carried out in the intermediate temperature region (1350~$\leq~T~\leq$~1400~$^{\circ}$C) can yield more than one crystallographic type of Ho$_{2}$Si$_{2}$O$_{7}$ (coexistence of B-type and C-type structures in the materials synthesised). Since holmium disilicate does not melt below $1750~^{\circ}$C~\cite{1970_Felsche}, it is expected that a crystal of Ho$_{2}$Si$_{2}$O$_{7}$ grown using the FZ method should crystallise in the structural type stable at the highest temperature i.e., E-type. To date, the magnetic properties of the various polymorphs of holmium disilicate have not been reported.

Thulium disilicate is dimorphic with temperature (see Table~\ref{Tab:R2Si2O7_temp_struct}), the structure stable at low temperature belongs to the B-type, whereas the high temperature polymorph is C-type~\cite{1970_Felsche}. Analogous to Er$_{2}$Si$_{2}$O$_{7}$ and Ho$_{2}$Si$_{2}$O$_{7}$, the melting point ($T_{m}~\approx~1700~^{\circ}$C) of the thulium disilicate is higher than the B~$\rightarrow$~C-type structural phase transition temperature ($T_{B~\rightarrow~C}~\approx~950~^{\circ}$C), one can presume that the crystallographic structure of a crystal of Tm$_{2}$Si$_{2}$O$_{7}$ grown using the FZ method will belong to the C-type polymorph. The magnetic properties of the high temperature C-type thulium disilicate have not yet been studied, nevertheless, due to the similarities between the arrangement of the Tm$^{3+}$ and Yb$^{3+}$ ions in the lattice, it is likely that Tm$_{2}$Si$_{2}$O$_{7}$ is likely to exhibit as interesting a magnetic behaviour as that observed previously in C-type Yb$_{2}$Si$_{2}$O$_{7}$~\cite{2019_Nair,2019_Hester,2020_Flynn}.

Crystals of the erbium disilicate have been grown previously using the FZ technique~\cite{2019_Nair}, however, in this study, the results of just one growth, under certain conditions, are presented. The holmium and thulium compounds have only been grown in crystal form using the flux method~\cite{1970_Felsche,1974_Wanklyn,1997_Maqsood,2000_Maqsood,2009_Maqsood,2014_Kahlenberg}. Firstly, we have optimised the crystal growth conditions for Er$_{2}$Si$_{2}$O$_{7}$, by performing a number of experiments and varying the growth parameters. We have also extended our study to the crystal growth of $R_{2}$Si$_{2}$O$_{7}$ (with \textit{R}~=~Ho, and Tm) and have successfully prepared, for the first time, using the optical FZ method, crystals of these rare-earth silicates. The crystals obtained are especially suitable for the investigation of the structural properties and magnetic behaviour of these materials. Our study shows that there is a direct correlation between the structural features and the magnetism exhibited.
\section{Experimental details}

The starting materials used for the synthesis of $R_{2}$Si$_{2}$O$_{7}$ (with \textit{R}~=~Er, Ho, and Tm) polycrystalline materials were rare-earth oxides, Er$_{2}$O$_{3}$, Ho$_{2}$O$_{3}$ and Tm$_{2}$O$_{3}$ (all of 99.9\% purity), and silica, SiO$_{2}$ (99.6\%). Crystals of rare-earth silicate compounds were then grown using a double ellipsoidal optical image furnace (NEC SC1MDH-11020, Canon Machinery Incorporated), equipped with two 1.5~kW halogen lamps.

The quality of the crystal boules was investigated using a Laue X-ray imaging system with a Photonic-Science Laue camera. Small quantities of each crystal were then ground and powder X-ray diffraction measurements were performed to determine the phase purity and to establish the crystallographic structure of the $R_{2}$Si$_{2}$O$_{7}$ crystals. It is essential to determine the structural type of each crystal. Room temperature diffractograms were collected on X-ray diffractometers (Panalytical and Bruker) using CuK$\alpha_{1}$ and CuK$\alpha_{2}$ radiation ($\lambda_{\mathrm{K}\alpha 1}~=~1.5406$~\AA~and $\lambda_{\mathrm{K}\alpha 2} ~=~1.5444$~\AA), over an angular range 10-70$^{\circ}$ or 10-90$^{\circ}$ in 2$\theta$, with a step size in the scattering angle 2$\theta$ of $0.013^{\circ}$ (Panalytical) and $0.016^{\circ}$ (Bruker). The analysis of the X-ray patterns was performed using the Fullprof software suite~\cite{1993_RodriguezCarvajal}.

Chemical composition analysis was carried out by energy dispersive x-ray spectroscopy (EDAX) using a scanning electron microscope on pieces cleaved from the $R_{2}$Si$_{2}$O$_{7}$ crystal boules.

Magnetic susceptibility measurements as a function of temperature were carried out on a ground piece of the Ho$_{2}$Si$_{2}$O$_{7}$ crystal down to 1.8 K in applied magnetic fields of 100 and 1000~Oe using a Quantum Design Magnetic Property Measurement System MPMS-5S superconducting quantum interference device (SQUID) magnetometer. 

Heat capacity measurements in zero applied magnetic field at temperatures from 0.5 to 300~K were carried out on Er$_{2}$Si$_{2}$O$_{7}$ and Tm$_{2}$Si$_{2}$O$_{7}$ crystals in a Quantum Design Physical Property Measurement System (PPMS) with a heat capacity option using a two-tau relaxation method.
\section{Results and discussion}
\subsection{Er$_{2}$Si$_{2}$O$_{7}$}

\subsubsection{Polycrystalline synthesis}

\begin{figure}[H]
\centering
\includegraphics[width=0.75\columnwidth]{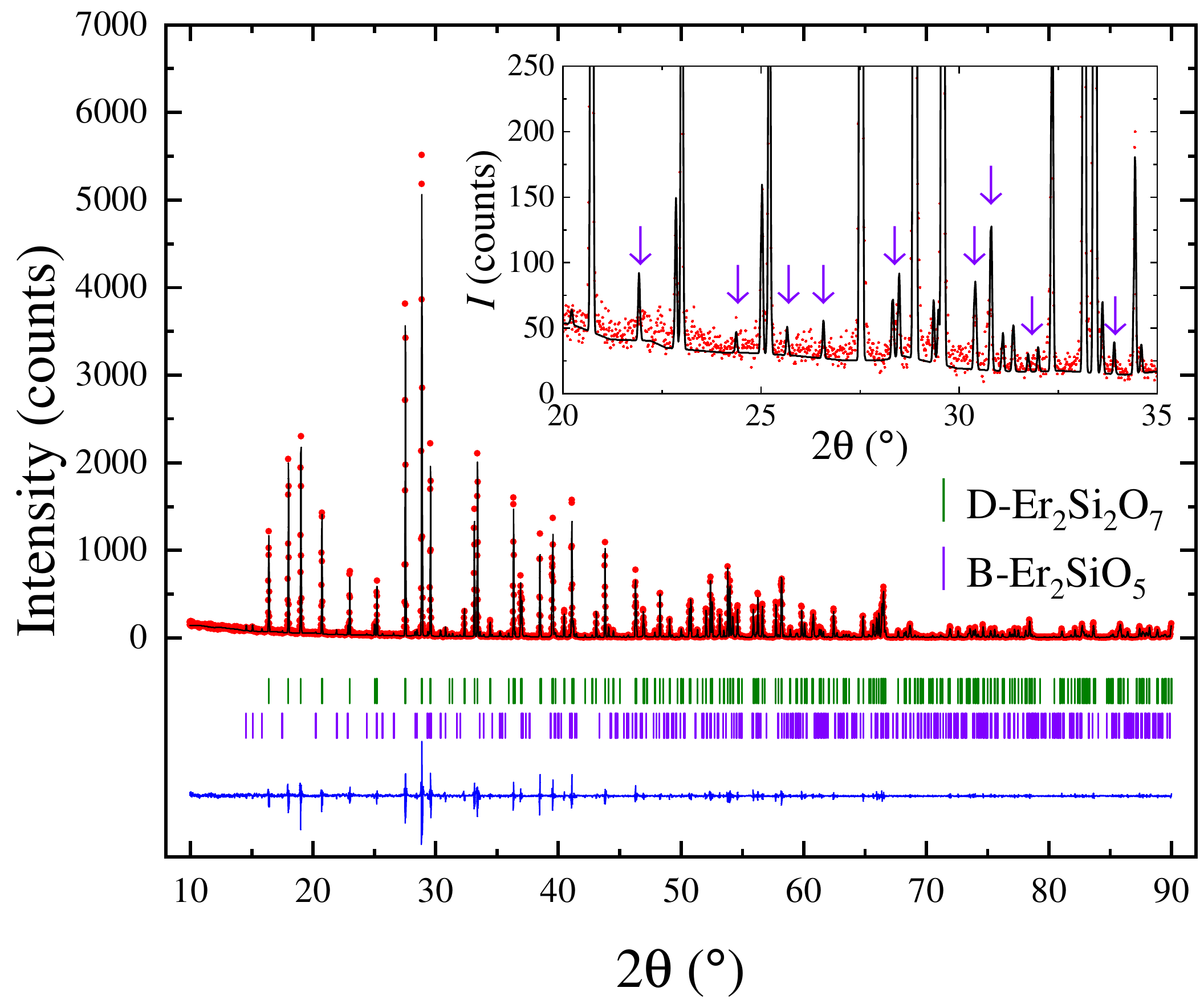}
\caption{Room temperature powder X-ray diffraction pattern of a Er$_{2}$Si$_{2}$O$_{7}$ polycrystalline sample (ESO). The experimental profile (red closed circles) and a full profile matching refinement (black solid line) made using the monoclinic ($P$2$_{1}$/$b$) D-type structure are shown, with the difference given by the blue solid line. The reflections of the D-type Er$_{2}$Si$_{2}$O$_{7}$ structure are indicated by green "\ding{120}", whereas the purple "\ding{120}"  show the reflections belonging to a B-type Er$_{2}$SiO$_{5}$ (monoclinic $C$2/$c$ structure) impurity. The inset shows the X-ray pattern in the range 20-35$^{\circ}$ scattering angle 2$\theta$, with the impurity reflections marked by purple arrows.}
\label{Fig:Er2Si2O7_polycrystal}
\end{figure}

D-type Er$_{2}$Si$_{2}$O$_{7}$ was first prepared in polycrystalline form by the conventional solid state synthesis method. Powders of Er$_{2}$O$_{3}$ and SiO$_{2}$ were weighed in stoichiometric amounts, mixed together and heat treated in air for several days (in 3 or 4~steps) at temperatures in the range 1400-1500~$^{\circ}$C. The annealed mixture (sample labelled ESO) was reground between each step of the synthesis to ensure good homogeneity and to facilitate the reaction of the starting materials. Nevertheless, even after 4 steps, each of long duration, a phase composition analysis by powder X-ray diffraction (goodness of fit (GOF)~=~1.68) reveals that a small quantity of a SiO$_{2}$ deficient erbium monosilicate impurity phase is present (see Fig.~\ref{Fig:Er2Si2O7_polycrystal}). This impurity, B-type Er$_{2}$SiO$_{5}$ (monoclinic structure, $C$2/$c$)~\cite{2008_Phanon}, persists even after the powder mixture is annealed at a higher temperature (1550-1600~$^{\circ}$C) and/or for extended periods of time (several days). The sintered material was isostatically pressed into rods (typically 6-8 mm diameter and 70-80 mm long) and sintered at 1500-1550~$^{\circ}$C in air for several days. The annealed rods were then used for the crystal growth.

\subsubsection{Crystal growth}

Crystals of erbium disilicate were grown using the FZ method, in static air atmosphere, at ambient pressure. In order to optimise the growth conditions, we have carried out a systematic study, by varying the growth speeds and the rotation rates of the two rods (feed and seed). The crystal growths were carried out at growth rates in the range 5-12~mm/h, and the feed and seed rods were counter-rotated, each at a rate of 10-25~rpm. Initially, a polycrystalline rod was used as a seed and once a good quality crystal boule was obtained, a crystal seed was used for subsequent growths. Er$_{2}$Si$_{2}$O$_{7}$ appears to melt congruently, and no deposition was observed on the quartz tube surrounding the feed and seed rods. The best quality crystals (assessment based on the analysis of the X-ray Laue diffraction) were obtained when growth rates of $\sim$ 10-12~mm/h were employed. The crystal quality appears to be independent of the rotation rate used for the seed rod.

\begin{figure}[H]
\centering
\includegraphics[width=0.75\columnwidth]{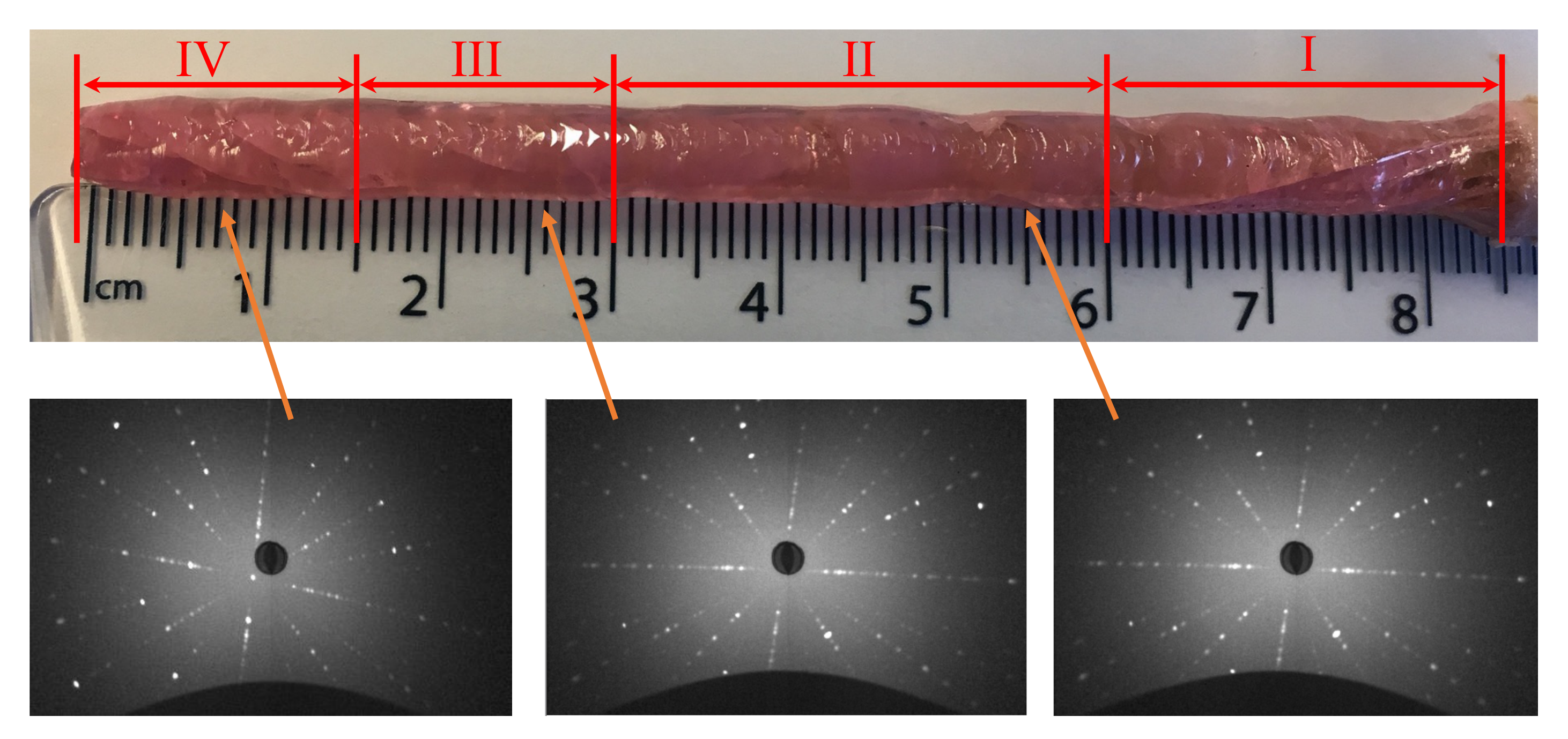}
\caption{Boule of Er$_{2}$Si$_{2}$O$_{7}$ prepared by the floating zone method in air atmosphere at a growth rate of 12~mm/h. Also shown are the X-ray Laue patterns, taken on one side of the boule at 3 points along its length. The crystal quality of each of the regions I-IV is discussed in the text.}
\label{Fig:Er2Si2O7_crystal}
\end{figure}

Er$_{2}$Si$_{2}$O$_{7}$ crystal boules prepared were typically 4-6~mm in diameter and 70-85~mm long. The boules tended to have thermally generated cracks in most cases, regardless of the growth rate employed. The crystals developed facets as they grew and two very strong facets were present on more than half the length of the grown crystals. All the erbium disilicate boules were a cloudy pink colour. The crystal boules of erbium disilicate are very fragile and all the crystals broke along the crystal growth axis into two long halves. Moreover, these two pieces cleaved a second time, perpendicular to the growth axis, thus forming crystal fragments of $\sim$ 5$\times$5$\times$3~mm$^{3}$. Figure~\ref{Fig:Er2Si2O7_crystal} shows a photograph of a crystal of Er$_{2}$Si$_{2}$O$_{7}$, which was grown using the optimal growth conditions, in air atmosphere, and at a growth speed of 12~mm/h.

X-ray Laue photographs taken of boules of Er$_{2}$Si$_{2}$O$_{7}$ confirm the good quality of the crystals. Typically, 2-3 grains are present and they extend along the length of each of the boules. A selection of Laue photographs taken along the length of an Er$_{2}$Si$_{2}$O$_{7}$ crystal boule is shown in Fig.~\ref{Fig:Er2Si2O7_crystal}. When a polycrystalline rod was used as a seed, the Laue patterns of the first 25 mm of the growth (region I) show that this region is either polycrystalline or poor quality crystal. We then note that identical Laue patterns were observed in region II for $\sim$ 30~mm, whereas our Laue examination of the boule at around 55~mm of growth, reveals the presence of 2 overlapping grains extending over a short length of $\sim$ 15~mm (region III) in the boule. For the remaining 15~mm of the boule (region IV), the Laue photographs show the existence of a single grain. The Laue patterns indicate that the $c^{*}$ axis direction is nearly orthogonal to the growth direction.

\begin{figure}[H]
\centering
\includegraphics[width=0.75\columnwidth]{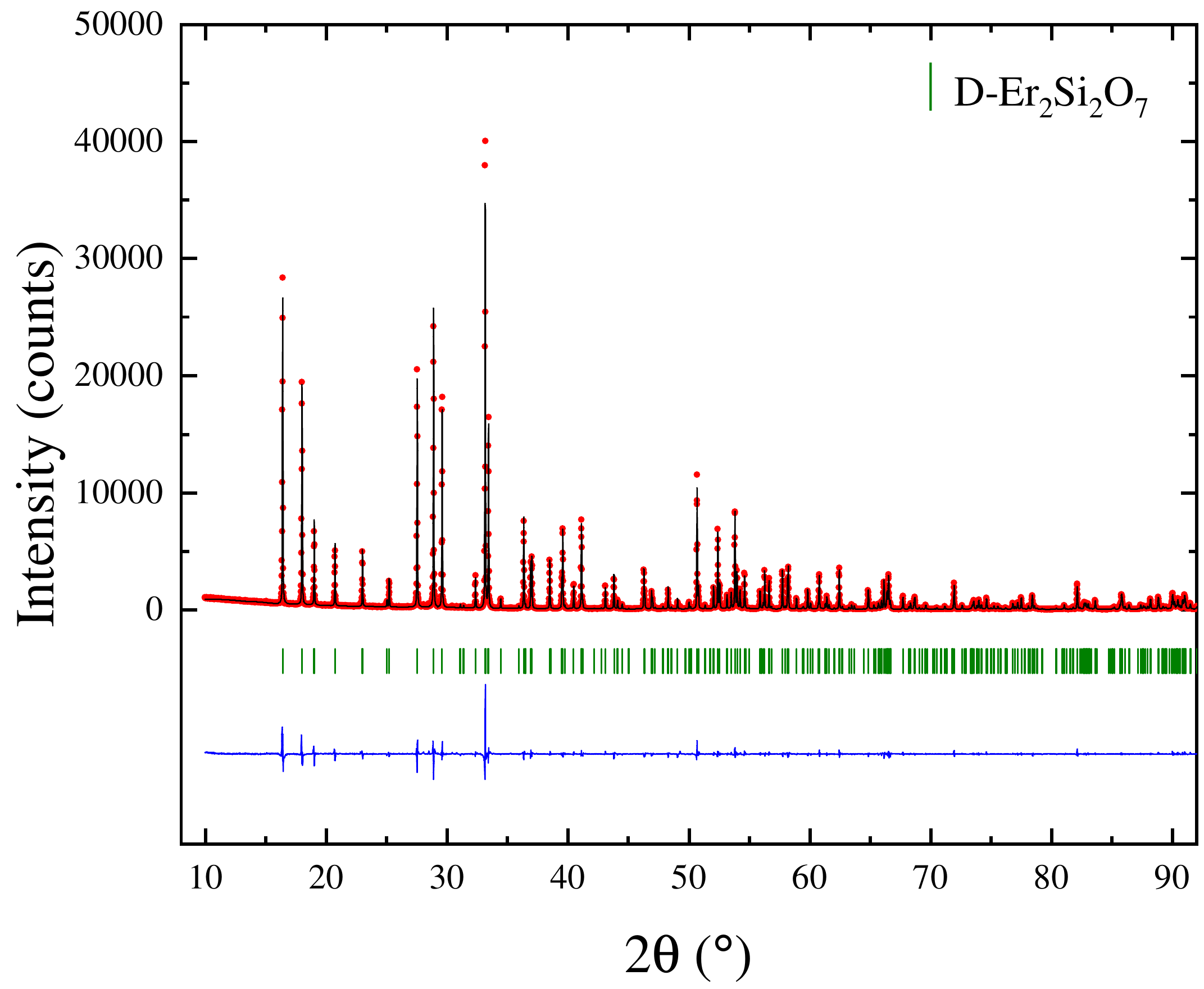}
\caption{Room temperature powder X-ray diffraction pattern of a ground Er$_{2}$Si$_{2}$O$_{7}$ crystal piece. The experimental profile (red closed circles) and a full profile matching refinement (black solid line) made using the D-type ($P$2$_{1}$/$b$) monoclinic structure are shown, with the difference given by the blue solid line.}
\label{Fig:Er2Si2O7_crystal_PXRD}
\end{figure}

Phase purity analysis was carried out on a ground crystal piece of Er$_{2}$Si$_{2}$O$_{7}$ and the powder X-ray diffraction pattern is shown in Fig.~\ref{Fig:Er2Si2O7_crystal_PXRD}. Profile matching (GOF~=~2.57) using the Fullprof software suite~\cite{1993_RodriguezCarvajal} indicates that the main phase is the monoclinic ($P$2$_{1}$/$b$) D-type Er$_{2}$Si$_{2}$O$_{7}$. There is no evidence that the monoclinic $C$2/$c$ B-type Er$_{2}$SiO$_{5}$ phase is present in the crystal, although it was observed as an impurity in the starting polycrystalline powders. The lattice parameters calculated from the profile matching were determined to be $a$~=~4.6908(2)~\AA, $b$~=~5.5615(2)~\AA\ and $c$~=~10.7991(2)~\AA, with the angle $\gamma$~=~96.040(2)$^{\circ}$. These are in reasonable agreement with previously published results~\cite{1970_Felsche}.

Composition analysis by EDAX was performed on a cleaved piece from the Er$_{2}$Si$_{2}$O$_{7}$ crystal boule. This showed that the cationic ratio averages of 1:1 for Er:Si for the bulk of the crystal. The average atomic percentages of Er, Si and O were 14.4(1)$\%$, 17.6(3)$\%$ and 68.0(3)$\%$ respectively. Given the limitations of this technique, the results are in reasonable agreement with the expected theoretical values of 18.2$\%$ for Er and Si, and 63.6$\%$ for O respectively.
\subsubsection{Heat capacity}

\begin{figure}[H]
\centering
\includegraphics[width=0.75\columnwidth]{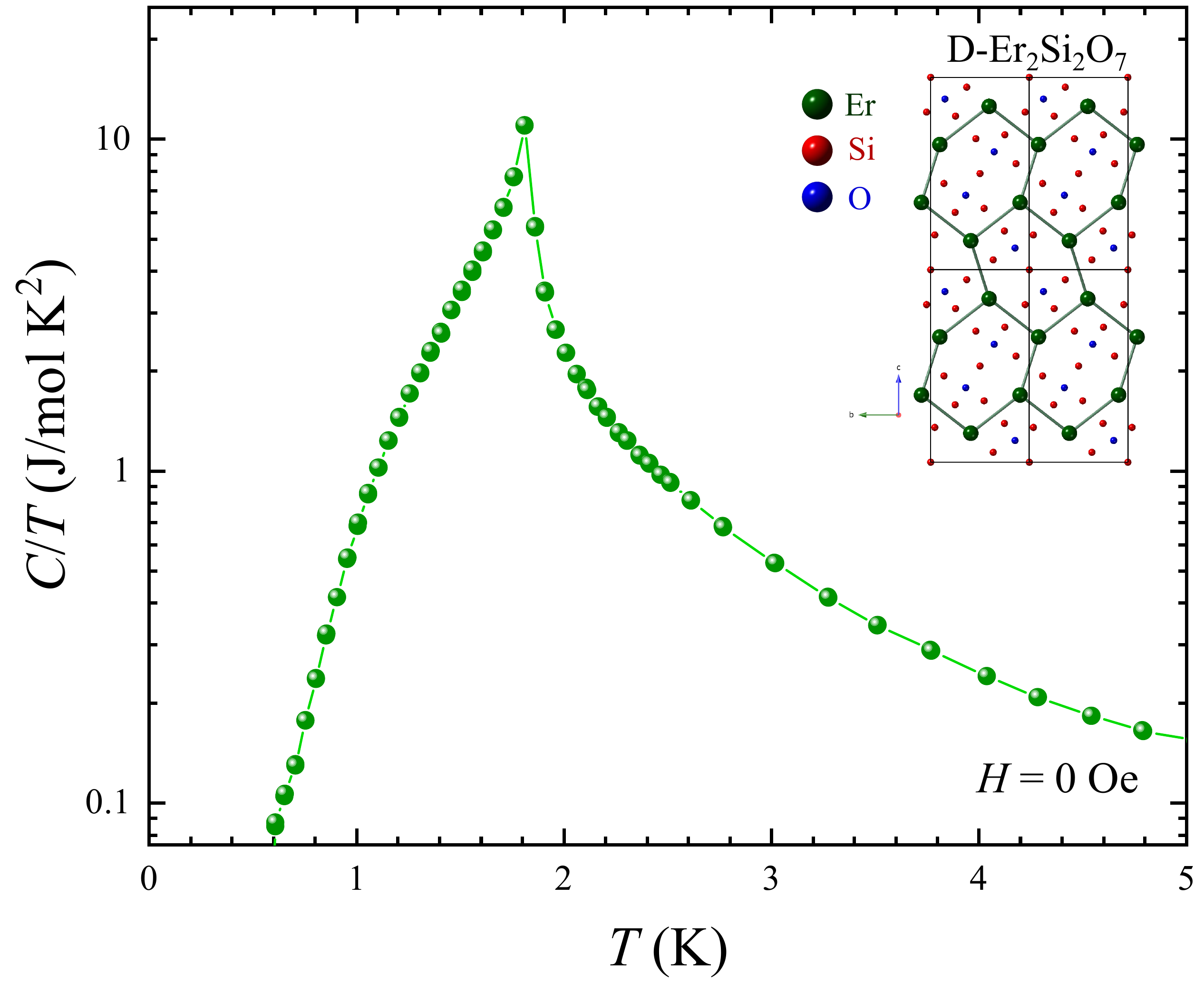}
\caption{Temperature dependence of $C/T$ for a D-type Er$_{2}$Si$_{2}$O$_{7}$ crystal piece in zero applied magnetic field. The inset shows the arrangement of the E$r^{3+}$ magnetic ions in the D-type crystallographic structure, emphasising the presence of distorted honeycombs.}
\label{Fig:Er2Si2O7_C/TvsT}
\end{figure}

D-type Er$_{2}$Si$_{2}$O$_{7}$ orders magnetically below 1.9(1)~K, with a presumed four-sublattice antiferromagnetic arrangement of Ising-like moments~\cite{1981_Maqsood,1986_Leask}. Heat capacity measurements performed in zero applied magnetic field on an unaligned fragment of an erbium disilicate crystal show a sharp peak in $C(T)/T$ at 1.80(2)~K (see Fig.~\ref{Fig:Er2Si2O7_C/TvsT}), confirming the ordering of the magnetic ions Er$^{3+}$. The temperature of the sharp feature observed in the heat capacity measurements further demonstrates that the FZ-grown Er$_{2}$Si$_{2}$O$_{7}$ crystal belongs to the structural type D, because C-type Er$_{2}$Si$_{2}$O$_{7}$ orders below $T_\mathrm{{N}}$~=~2.50(5)~K~\cite{1979_Maqsood,2000_Maqsood}. The results are in agreement with the previous results published on a Er$_{2}$Si$_{2}$O$_{7}$ crystal grown using the FZ method~\cite{2019_Nair}. Extensive studies of the magnetic ground state of D-type Er$_{2}$Si$_{2}$O$_{7}$ are described elsewhere in a more detailed paper~\cite{2020_Petrenko}.

\subsection{Ho$_{2}$Si$_{2}$O$_{7}$}

\subsubsection{Polycrystalline synthesis}

The two structural types of holmium disilicate that are promising from a magnetic point of view are the C and D polymorphs. The magnetic ions are arranged in these crystallographic structures, in such a way that they form a distorted honeycomb-like lattice (see Figs.~\ref{Fig:R2Si2O7_structure}(b-c)).

Ho$_{2}$Si$_{2}$O$_{7}$ was first prepared in polycrystalline form by the conventional solid state synthesis method. Due to the fact that the temperatures of the structural phase transitions are close to one another~\cite{1970_Felsche,2009_Maqsood}, we have carried out several synthesis attempts in order to determine the optimal conditions for preparing C or D-type holmium disilicate compounds.

\begin{figure}[H]
\centering
\includegraphics[width=0.75\columnwidth]{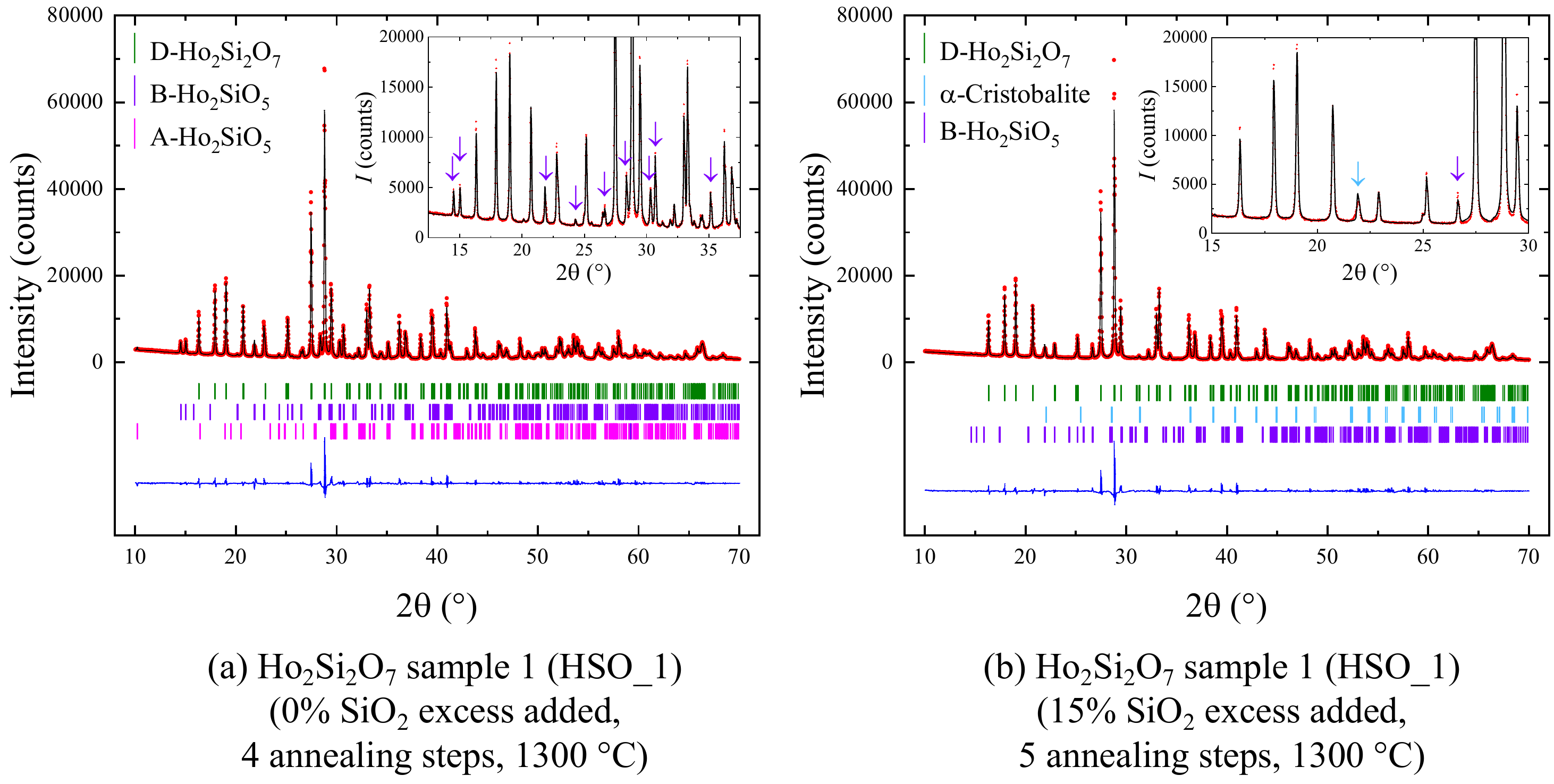} \\
\includegraphics[width=0.75\columnwidth]{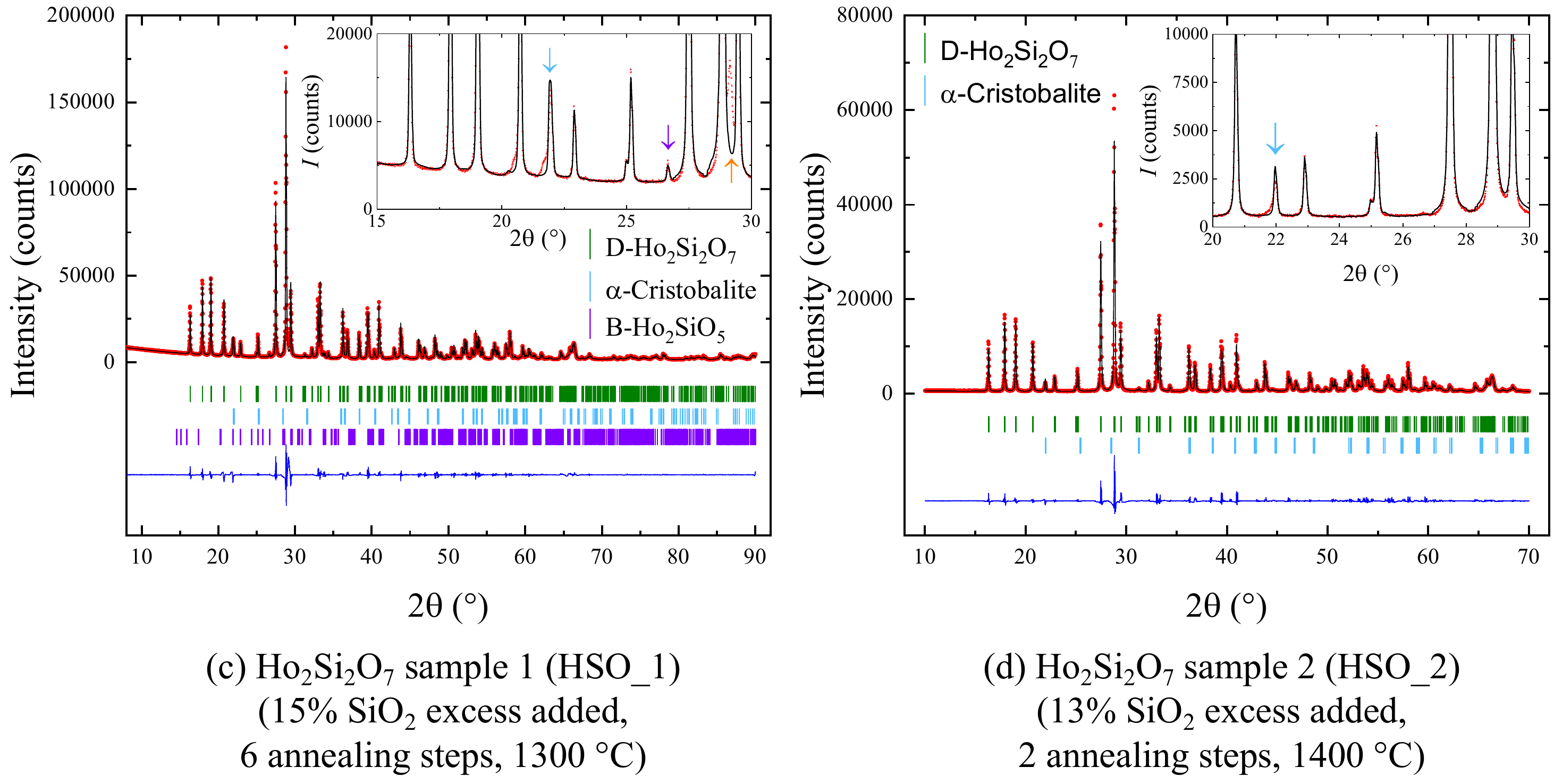} \\
\includegraphics[width=0.75\columnwidth]{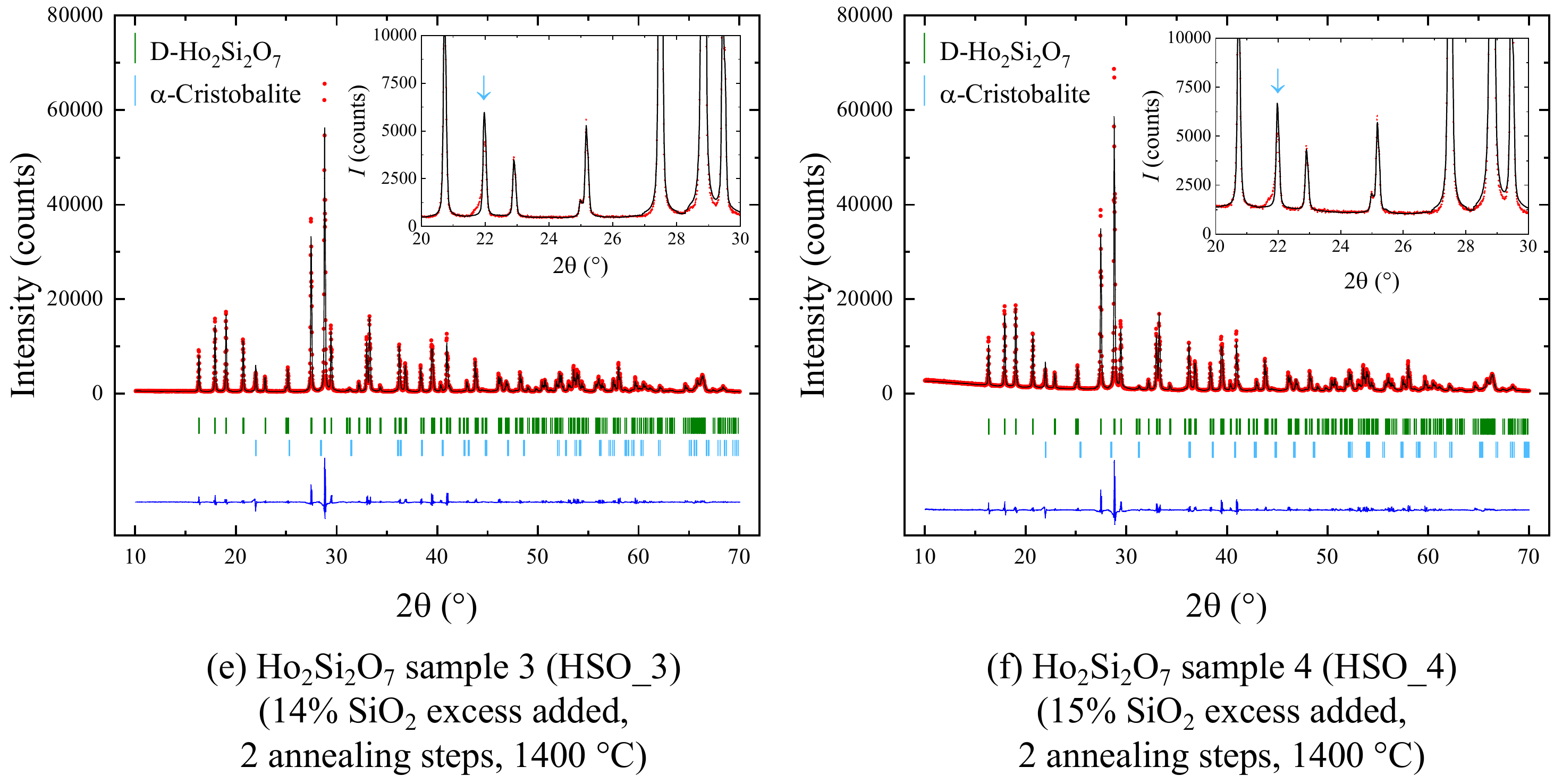}
\caption{Room temperature powder X-ray diffraction pattern of four Ho$_{2}$Si$_{2}$O$_{7}$ polycrystalline samples (HSO\_1, HSO\_2, HSO\_3 and HSO\_4) made using different amounts of excess SiO$_{2}$. The experimental profile (red closed circles) and a full profile matching refinement (black solid line) made using the monoclinic ($P$2$_{1}$/$b$) D-type structure are shown, with the difference given by the blue solid line. The reflections of the D-type Ho$_{2}$Si$_{2}$O$_{7}$ structure are indicated by green "\ding{120}"; purple "\ding{120}"  show the reflections belonging to a B-type Ho$_{2}$SiO$_{5}$ (monoclinic $C$2/$c$ structure) impurity; pink "\ding{120}" mark the reflections belonging to A-type Ho$_{2}$SiO$_{5}$ (monoclinic $P$2$_{1}$/$c$ structure ) impurity; light blue "\ding{120}" identify the Bragg peaks belonging to unreacted SiO$_{2}$ crystallising in a tetragonal ($P$4$_{1}$2$_{1}$2) structure ($\alpha$-cristobalite). The insets show the X-ray patterns over a reduced range of the scattering angle 2$\theta$, with the impurity reflections marked by coloured arrows. In (c) an orange arrow marks the reflection belonging to an impurity phase that could not be identified.}
\label{Fig:Ho2Si2O7_polycrystal_PXRD}
\end{figure}

In a first attempt (sample labelled HSO\_1), stoichiometric amounts of Ho$_{2}$O$_{3}$ and SiO$_{2}$ powders were ground together and reacted in air for several days (in 4~steps) at 1300~$^{\circ}$C, with intermediate grindings. Analysis (GOF~=~3.81) of the X-ray diffraction pattern collected at room temperature on this powder (see Fig.~\ref{Fig:Ho2Si2O7_polycrystal_PXRD}(a)) indicates that the main phase is the D-type Ho$_{2}$Si$_{2}$O$_{7}$. Nevertheless, there are several Bragg peaks which could not be indexed with the monoclinic ($P$2$_{1}$/$b$) space group, and arise due to the presence of a holmium monosilicate impurity, Ho$_{2}$SiO$_{5}$. This impurity phase is present in two structural types, a monoclinic ($P$2$_{1}$/$c$) structure (A-type), and a monoclinic ($C$2/$c$) arrangement (B-type)~\cite{1984_Maqsood}.

The difficulty in stabilising the Si-rich phase, Ho$_{2}$Si$_{2}$O$_{7}$, is probably due to the reactivity of the starting silica powder. Felsche previously reported the synthesis of rare-earth disilicates starting with the highly reactive form of SiO$_{2}$, cristobalite, a very high temperature polymorph of quartz~\cite{1970_Felsche}. In all our synthesis experiments, we have used a less reactive form of silica. To overcome this drawback, one option is to pre-anneal the starting reagent, silica, in air at high temperatures (1500-1650~$^{\circ}$C), however, this process does not produce a direct conversion to cristobalite~\cite{1961_Chaklader}. We have thus opted for a second solution, i.e. the synthesis using an excess of SiO$_{2}$.

An excess of 15\% silica was added to the reacted stoichiometric powder obtained previously. The powder mixture was then heated in air for several days (in 1~step) at 1300~$^{\circ}$C. Phase purity analysis was carried out and the diffractogram is shown in Fig.~\ref{Fig:Ho2Si2O7_polycrystal_PXRD}(b). Profile matching (GOF~=~3.71) indicates that the main phase is D-type Ho$_{2}$Si$_{2}$O$_{7}$ and that there are two impurity phases present, unreacted SiO$_{2}$ crystallising in a tetragonal ($P$4$_{1}$2$_{1}$2) structure ($\alpha$-cristobalite~\cite{1994_Downs}) and B-type Ho$_{2}$SiO$_{5}$. To ensure good homogeneity and to facilitate the reaction, another annealing was carried out for several days at 1300~$^{\circ}$C. Room temperature powder X-ray diffraction measurements were again performed to determine the phase purity of the polycrystalline material. The X-ray diffraction pattern (see Fig.~\ref{Fig:Ho2Si2O7_polycrystal_PXRD}(c)) (GOF~=~6.46) reveals the presence of D-type Ho$_{2}$Si$_{2}$O$_{7}$, $\alpha$-cristobalite and B-type Ho$_{2}$SiO$_{5}$. In addition, a fourth chemical phase is present (indicated by the existence of one unindexed Bragg peak at $\sim$~29.1$^{\circ}$ in 2$\theta$), however, this impurity could not be identified, due to the reduced intensity of the peak. Two additional shoulders can be observed at $\sim$~20.5 and 21.7$^{\circ}$ in 2$\theta$, however, due to their extremely reduced intensities, their presence could not be correlated with any known chemical phase. Additional trials were therefore performed in order to optimise the synthesis conditions and obtain pure phase D-type Ho$_{2}$Si$_{2}$O$_{7}$ polycrystalline material.

In a second attempt, we have prepared three samples (labelled HSO\_2, HSO\_2 and HSO\_3) starting with stoichiometric amounts of Ho$_{2}$O$_{3}$ and different amounts of excess SiO$_{2}$ (13, 14 and 15\%). The starting oxides were mixed together and heat treated in air for several days (in 2~steps) at 1400~$^{\circ}$C, with an intermediate grinding. The powder X-ray diffraction patterns obtained for the holmium disilicate powders prepared using excess silica are shown in Figs.~\ref{Fig:Ho2Si2O7_polycrystal_PXRD}(d)-(f). An analysis of the patterns provided a good fit (GOF~=~3.71, 3.70 and 3.75 for 13, 14 and 15\% excess SiO$_{2}$ respectively) to the D-type Ho$_{2}$Si$_{2}$O$_{7}$. Additionally, in all three patterns, there is one Bragg peak which could not be indexed with the monoclinic ($P$2$_{1}$/$b$) space group. This Bragg peak is attributed to the presence of a small amount of unreacted SiO$_{2}$, in the form of $\alpha$-cristobalite. In addition, the existence of an unindexed shoulder at $\sim$~21.7$^{\circ}$ in 2$\theta$, suggests the presence of a third chemical phase, however this impurity could not be identified, due to the reduced intensity of the peak. The best results in the synthesis of D-type Ho$_{2}$Si$_{2}$O$_{7}$ powders were obtained when an excess of 13\% silica was used (see Fig.~\ref{Fig:Ho2Si2O7_polycrystal_PXRD}(d)). 

Throughout the synthesis procedure, extra care has been taken to minimise the amount of unreacted SiO$_{2}$ in the polycrystalline material used to prepare the feed and seed rods. To ensure a higher reactivity of the starting silica reagent, silica was pre-annealed in air at 1400~$^{\circ}$C for 24~h. To ensure that the chemical reaction is complete, we have started with a slightly larger amount of excess (14\%) than determined previously to be optimal and we have reacted the powder mixtures in air for several days (in 3 steps). The sintered material was isostatically pressed into rods (typically 6-8 mm diameter and 70-80 mm long) and sintered at 1400-1450~$^{\circ}$C in air for several days. The annealed rods were then used for the crystal growth.

\subsubsection{Crystal growth}

We have first grown crystal boules of Ho$_{2}$Si$_{2}$O$_{7}$ using the FZ method, in static air atmosphere, at ambient pressure. The crystal growths were carried out at growth rates in the range 5-15~mm/h, and the feed and seed rods were counter-rotated, each at a rate of 15-25~rpm. Polycrystalline rods were used as seed rods. Ho$_{2}$Si$_{2}$O$_{7}$ appears to melt congruently, and no deposition was observed on the quartz tube surrounding the feed and seed rods. The crystals did not develop any facets as they grew and the crystal boules of holmium disilicate were very fragile. X-ray Laue photographs taken of these boules of Ho$_{2}$Si$_{2}$O$_{7}$ reveal a poor crystalline quality of the crystal boules.

\begin{figure}[H]
\centering
\includegraphics[width=0.75\columnwidth]{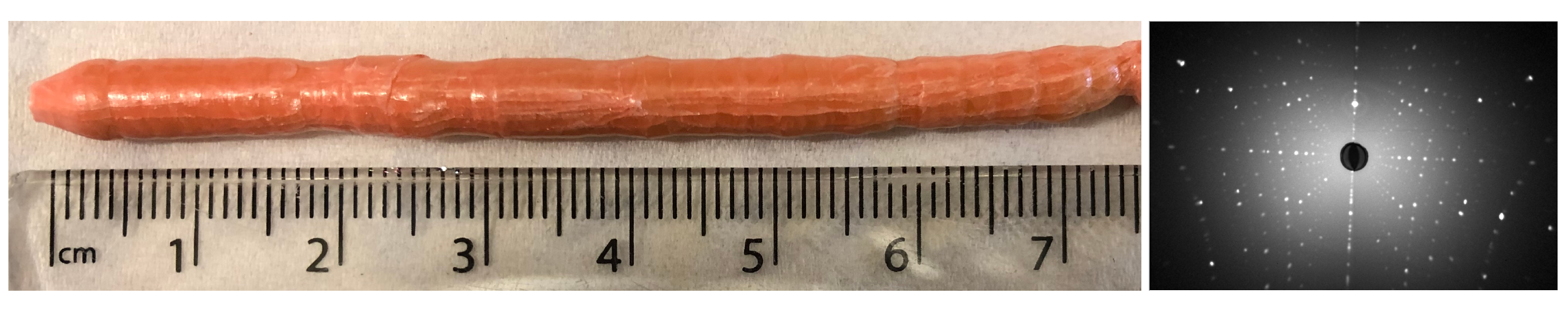}
\caption{Boule of Ho$_{2}$Si$_{2}$O$_{7}$ prepared by the floating zone method in air atmosphere, at pressures in the range 1-2 bars, in a flow of air of 0.1-0.2~L/min, using a growth speed of 8~mm/h. Also shown is the X-ray Laue pattern of one of the sides of the Ho$_{2}$Si$_{2}$O$_{7}$ crystal boule. }
\label{Fig:Ho2Si2O7_crystal}
\end{figure}

The crystal growth of Ho$_{2}$Si$_{2}$O$_{7}$ was also carried out in air at pressures in the range 1-2 bars, a flow of air of 0.1-0.2~L/min. A growth rate of 8~mm/h was used, with the feed and seed rods counter-rotating at a rate of 10 and 25~rpm, respectively. A boule prepared in static air atmosphere, at a growth rate of 15~mm/h was used as a seed for this crystal growth. The Ho$_{2}$Si$_{2}$O$_{7}$ crystal boule obtained was 5~mm in diameter and 75~mm long. The boule tended to have thermally generated cracks, and it developed facets as it grew and two very strong facets were present on more than half the length of the grown boule. All the holmium disilicate crystal boules obtained were a pale orange colour. Figure~\ref{Fig:Ho2Si2O7_crystal} shows a photograph of a crystal boule of Ho$_{2}$Si$_{2}$O$_{7}$, grown in air atmosphere, at pressures in the range 1-2 bars, in a flow of air of 0.1-0.2~L/min, using a growth speed of 8~mm/h. X-ray Laue photographs taken of this boule of Ho$_{2}$Si$_{2}$O$_{7}$ show that the crystal boule consists of a collection of several grains. Nevertheless, long needle-like single crystals, $\sim$ 20$\times$2$\times$1-2~mm$^{3}$, could be isolated from the crystal boule. The size of these crystals makes them suitable for the study of the magnetic behaviour of this system. 

\begin{figure}[H]
\centering
\includegraphics[width=0.75\columnwidth]{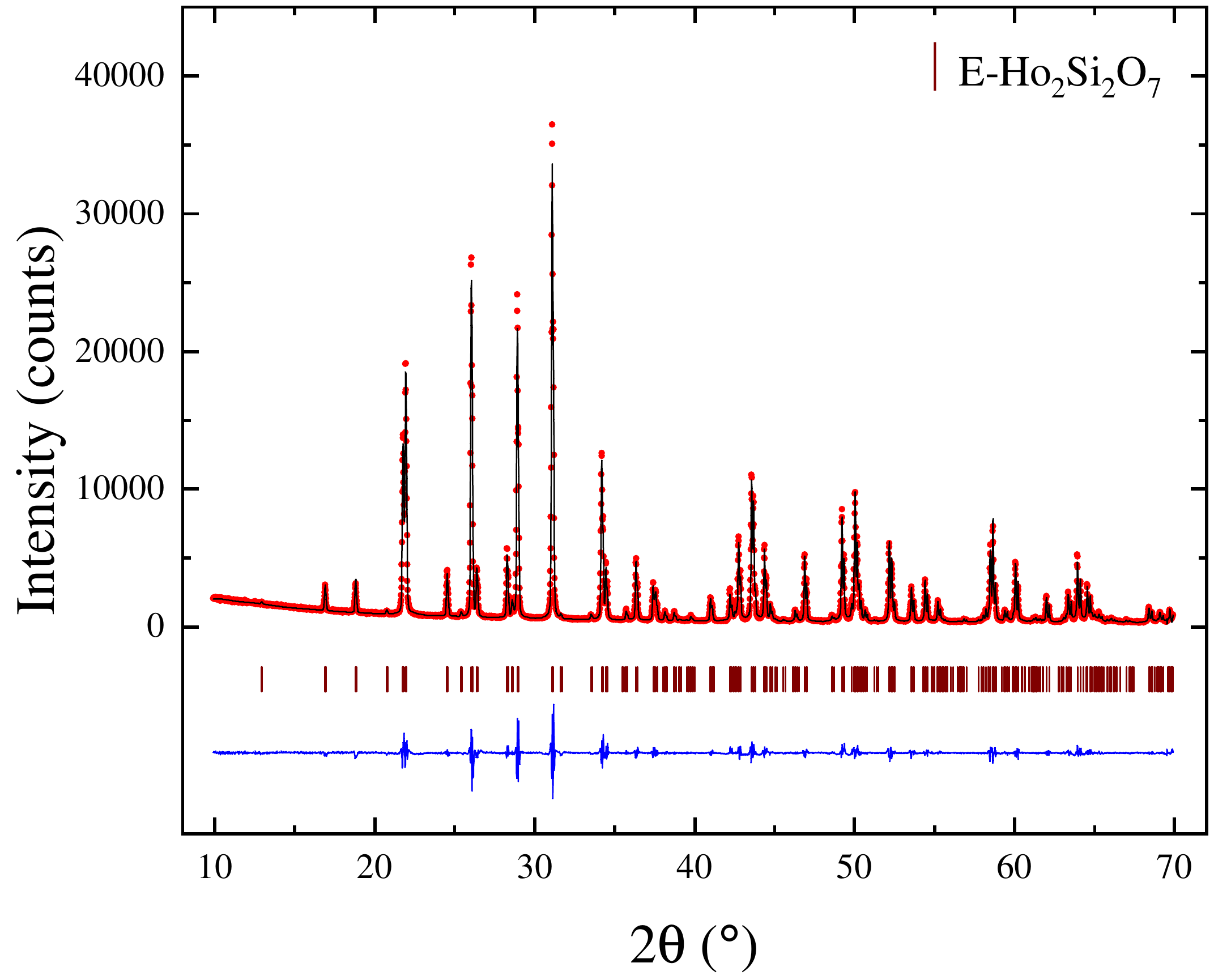}
\caption{Room temperature powder X-ray diffraction pattern of a ground piece of Ho$_{2}$Si$_{2}$O$_{7}$ crystal. The experimental profile (red closed circles) and a full profile matching refinement (black solid line) made using the E-type ($Pna2_{1}$) orthorhombic structure are shown, with the difference given by the blue solid line.}
\label{Fig:Ho2Si2O7_crystal_PXRD}
\end{figure}

Holmium disilicate does not melt below $1750~^{\circ}$C, and the D~$\rightarrow$~E-type structural phase transition occurs at a lower temperature than the melting point~\cite{1970_Felsche}. One thus expects a crystal of Ho$_{2}$Si$_{2}$O$_{7}$ grown using the FZ method to crystallise in the structural type stable at the highest temperature ($T~\geq$~1500~$^{\circ}$C) i.e., E-type. Phase purity analysis of a ground piece of the holmium disilicate crystal boule grown in air atmosphere, at pressures in the range 1-2 bars, a flow of air of 0.1-0.2~L/min, using a growth rate of 8~mm/h (see Fig.~\ref{Fig:Ho2Si2O7_crystal_PXRD}), shows that the main phase is E-type Ho$_{2}$Si$_{2}$O$_{7}$ (orthorhombic $Pna2_{1}$ structure), with no significant impurity phases present. Profile matching (GOF = 2.98) was carried out and the lattice parameters were determined to be $a$~=~13.6770(3)~\AA, $b$~=~5.0235(3)~\AA\ and $c$~=~8.1598(3)~\AA. These are slightly smaller than the previously published results on flux grown crystals of E-type Ho$_{2}$Si$_{2}$O$_{7}$~\cite{1970_Felsche,2009_Maqsood}.

Composition analysis by EDAX was performed on a cleaved piece from the Ho$_{2}$Si$_{2}$O$_{7}$ crystal boule. Given the limitations of this analysis method, the average atomic percentages of 16.0(6)$\%$, 16.6(2)$\%$ and 67.4(6)$\%$ for Ho, Si and O, respectively, are in reasonable agreement with the expected theoretical values (18.2$\%$ for Ho and Si, and 63.6$\%$ for O) for the Ho$_{2}$Si$_{2}$O$_{7}$ phase.

\subsubsection{Magnetisation}

\begin{figure}[H]
\centering
\includegraphics[width=0.75\columnwidth]{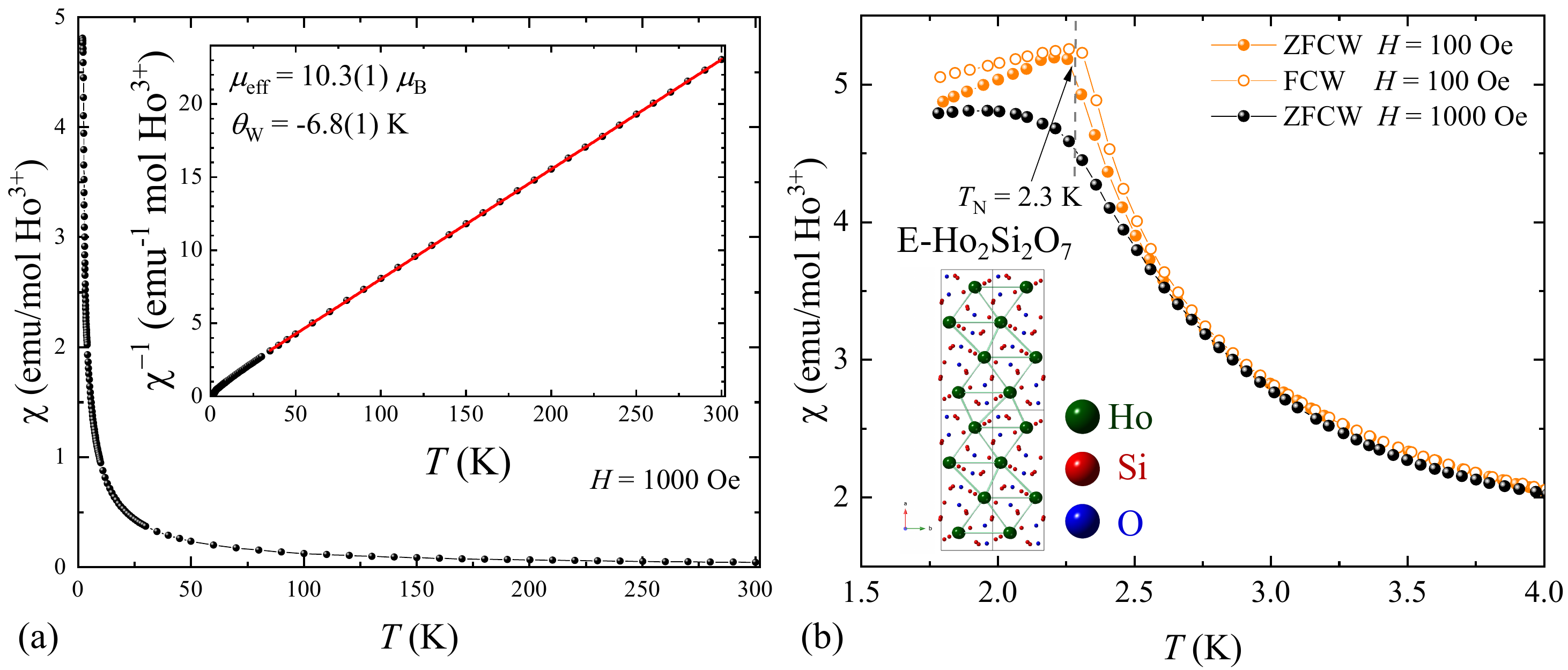}
\caption{(a) Temperature dependence of the \textit{dc} magnetic susceptibility, $\chi$, in the range 1.8 to 300~K for an E-type Ho$_{2}$Si$_{2}$O$_{7}$ ground crystal sample in an applied magnetic field of 1000~Oe. The inset shows $\chi^{-1}$ versus $T$ and the fit using the Curie-Weiss law to the data in the temperature range 35 to 300~K. (b) Magnetic susceptibility data in a reduced temperature range (1.8-4~K) in applied magnetic fields of 100 and 1000~Oe. The inset shows the arrangement of the Ho$^{3+}$ magnetic ions in the E-type crystallographic structure.}
\label{Fig:Ho2Si2O7_Magn}
\end{figure}

Zero-field-cooled-warming (ZFCW) and field-cooled-warming (FCW) magnetisation versus temperature curves were measured on a ground piece of the E-type Ho$_{2}$Si$_{2}$O$_{7}$. The temperature dependence of the \textit{dc} magnetic susceptibility, $\chi\left(T\right)$, and reciprocal \textit{dc} magnetic susceptibility $\chi^{-1}\left(T\right)$ are shown in Fig.~\ref{Fig:Ho2Si2O7_Magn}. The magnetic susceptibility in an applied magnetic field of 1000~Oe exhibits a monotonic increase when cooling from 300 to 1.8~K, and an anomaly is observed at low temperature. A fit of the $\chi^{-1}\left(T\right)$ data to a Curie-Weiss law over an extended temperature range (35-300~K) (see Fig.~\ref{Fig:Ho2Si2O7_Magn}(a) inset) shows that for 35~$\leq$~$T$~$\leq$~300~K, E-type Ho$_{2}$Si$_{2}$O$_{7}$  has an effective moment of $\mu_\mathrm{{eff}}~=~10.3(1)\mu_\mathrm{{B}}$ and a Weiss temperature of $\theta_\mathrm{{W}}~=~-6.8(1)$~K. The effective moment of Ho$^{3+}$ in holmium disilicate is in agreement with the magnetic moment of a free Ho$^{3+}$ in the ground state $^{5}I_{8}$. The $\theta_\mathrm{{W}}$ value indicates an antiferromagnetic coupling between the Ho spins.

A measurement of the temperature dependence of the susceptibility was also performed in an applied magnetic field of 100~Oe. The examination of the magnetisation at low temperatures (1.8~$<$~$T$~$<$~4~K) reveals a bifurcation of the ZFCW and FCW susceptibility curves below $\sim$~2.55(5)~K, shown in Fig.~\ref{Fig:Ho2Si2O7_Magn}(b). The feature centred around 2.30(5)~K suggests the antiferromagnetic ordering between the Ho$^{3+}$ ions. In order to establish the magnetic structure of E-type Ho$_{2}$Si$_{2}$O$_{7}$ and its evolution with the applied magnetic field, detailed studies of the magnetic behaviour of this system are in progress.

\subsection{Tm$_{2}$Si$_{2}$O$_{7}$}

\subsubsection{Polycrystalline synthesis}
The Tm$_{2}$Si$_{2}$O$_{7}$ polymorph stable at high temperature is C-type (see Fig.~\ref{Fig:R2Si2O7_structure}(b)). The arrangement of the magnetic ions in the crystallographic structure is similar to what is observed in Yb$_{2}$Si$_{2}$O$_{7}$, and this motivated us to embark on the study of C-type thulium disilicate.

\begin{figure}[H]
\centering
\includegraphics[width=0.75\columnwidth]{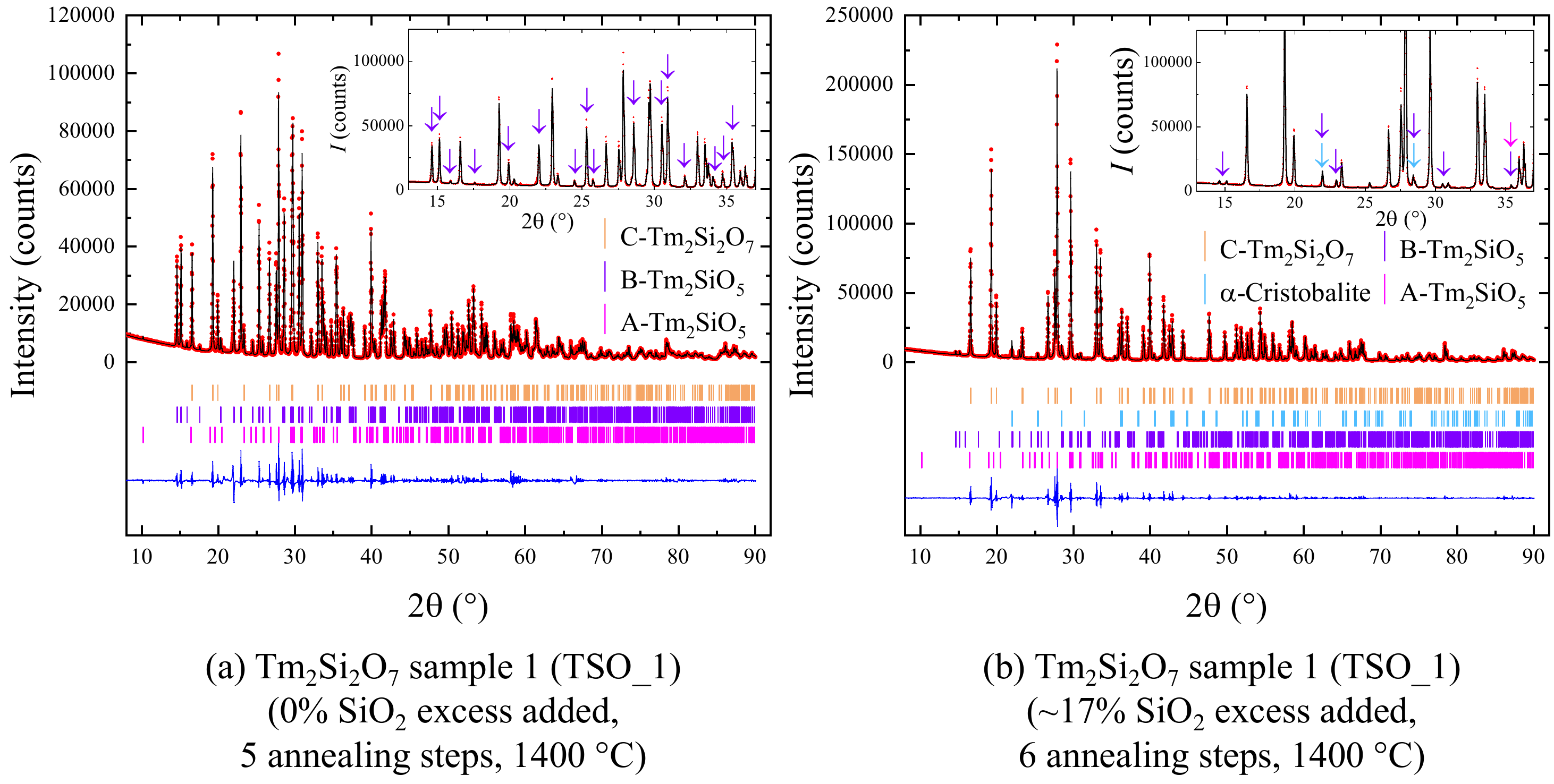} \\
\includegraphics[width=0.75\columnwidth]{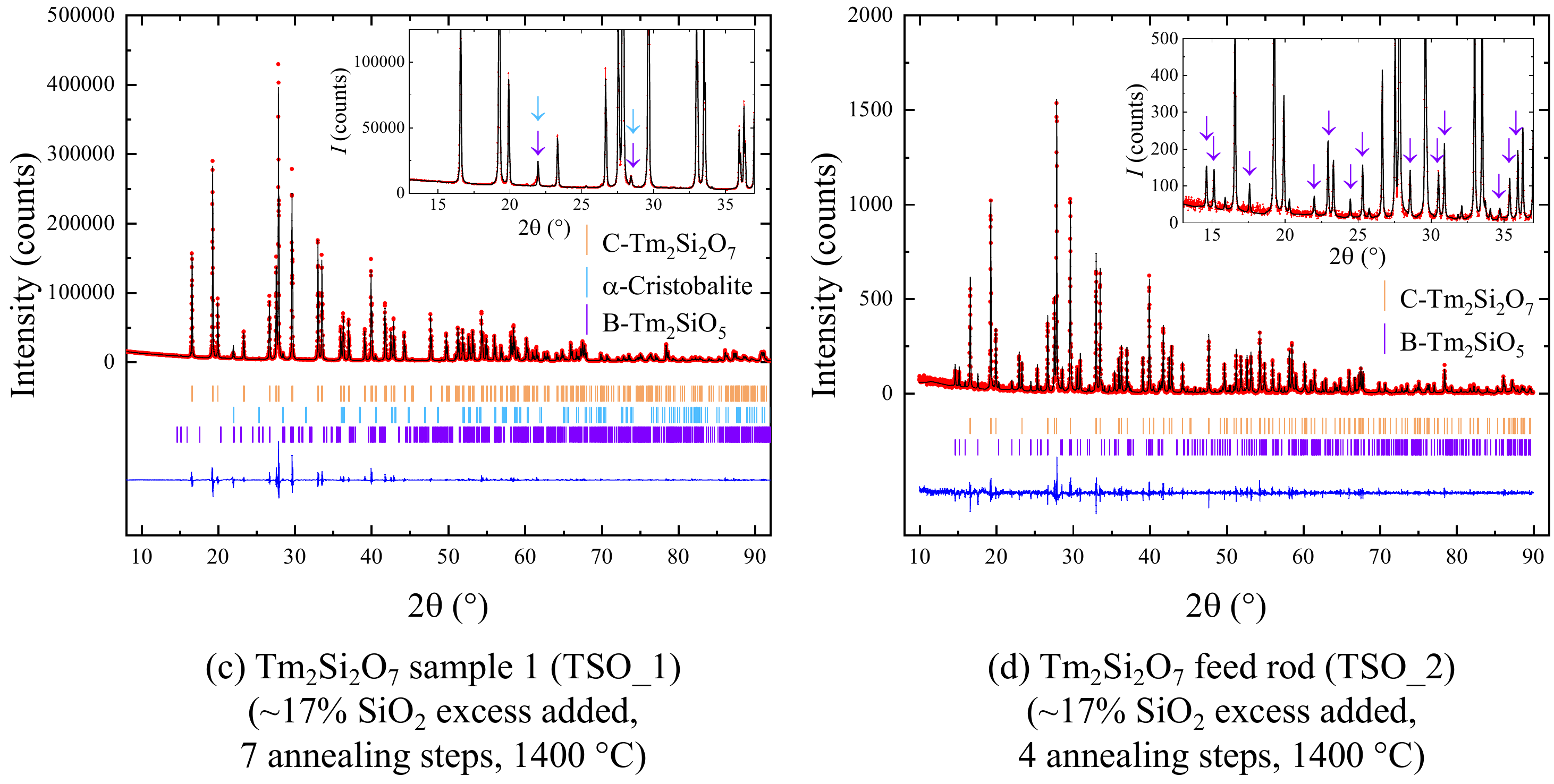}
\caption{Room temperature powder X-ray diffraction pattern of two Tm$_{2}$Si$_{2}$O$_{7}$ polycrystalline samples (TSO\_1 and TSO\_2) made with and without excess SiO$_{2}$. The experimental profile (red closed circles) and a full profile matching refinement (black solid line) made using the monoclinic ($C$2/$m$) C-type structure are shown, with the difference given by the blue solid line. The reflections of the C-type Tm$_{2}$Si$_{2}$O$_{7}$ structure are indicated by orange "\ding{120}"; purple "\ding{120}"  show the reflections belonging to a B-type Tm$_{2}$SiO$_{5}$ (monoclinic $C$2/$c$ structure) impurity; pink "\ding{120}" mark the reflections belonging to A-type Tm$_{2}$SiO$_{5}$ (monoclinic $P$2$_{1}$/$c$ structure ) impurity; light blue "\ding{120}" identify the Bragg peaks belonging to unreacted SiO$_{2}$ crystallising in a tetragonal ($P$4$_{1}$2$_{1}$2) structure ($\alpha$-cristobalite). The insets show the X-ray patterns over a reduced range of the scattering angle 2$\theta$, with the impurity reflections marked by coloured arrows.}
\label{Fig:Tm2Si2O7_polycrystal_PXRD}
\end{figure}

To prepare C-type Tm$_{2}$Si$_{2}$O$_{7}$ powders, stoichiometric amounts of Tm$_{2}$O$_{3}$ and SiO$_{2}$ were thoroughly ground, pressed into pellets and then heated several times in air, for several days, at 1400~$^{\circ}$C, above the B~$\rightarrow$~C-type structural phase transition temperature~\cite{1970_Felsche}, in 5~steps, with intermediate grinding. A powder X-ray diffraction measurement was carried out at room temperature in order to check the composition of the thulium disilicate polycrystalline sample prepared (labelled TSO\_1). Analysis of the pattern (GOF~=~7.26) using the FullProf software suite indicates that although the main phase is the monoclinic ($C$2/$m$) C-type structure of Tm$_{2}$Si$_{2}$O$_{7}$, there are several peaks belonging to a thulium monosilicate impurity, Tm$_{2}$SiO$_{5}$ (see Fig.~\ref{Fig:Tm2Si2O7_polycrystal_PXRD}(a)). This impurity is present in two structural phases, a monoclinic ($P$2$_{1}$/$c$) structure (A-type), and a monoclinic ($C$2/$c$) arrangement (B-type)~\cite{2001_Tian,2016_Tian2}. 

To obtain single phase polycrystalline material of C-type Tm$_{2}$Si$_{2}$O$_{7}$, we have adopted a similar approach to the one used for the synthesis of Ho$_{2}$Si$_{2}$O$_{7}$. An excess ($\sim$ 17\%) of SiO$_{2}$ was added to the pre-reacted stoichiometric powder obtained. The powder mixture was then pressed into pellets and heated in air for several days (in 1~step) at 1400~$^{\circ}$C. The X-ray diffraction pattern (GOF~=~7.45) collected on this powder shows that the polycrystalline material is a mixture of phases (see Fig.~\ref{Fig:Tm2Si2O7_polycrystal_PXRD}(b)). The main phase is C-type Tm$_{2}$Si$_{2}$O$_{7}$, but there are several impurity peaks belonging to A and B-type Tm$_{2}$SiO$_{5}$, as well as unreacted SiO$_{2}$ crystallised in the $\alpha$-cristobalite tetragonal ($P$4$_{1}$2$_{1}$2) structure. The powder mixture was pelletised again and reacted in air for several days at 1400~$^{\circ}$C. Phase purity analysis of the resulting powder by X-ray diffraction (GOF~=~7.73) shows the existence of three phases, C-type Tm$_{2}$Si$_{2}$O$_{7}$, B-type Tm$_{2}$SiO$_{5}$ and $\alpha$-cristobalite. Figures~\ref{Fig:Tm2Si2O7_polycrystal_PXRD}(a-c) show that the amount of the Tm$_{2}$SiO$_{5}$ impurity is significantly reduced due to the addition of SiO$_{2}$.

A larger amount of polycrystalline sample of C-type Tm$_{2}$Si$_{2}$O$_{7}$ was prepared using a stoichiometric amount of Tm$_{2}$O$_{3}$ and $\sim$ 17\% excess SiO$_{2}$. The starting materials were mixed together and heat treated in air for several days (in 3~steps) at 1400~$^{\circ}$C, with an intermediate grinding. The sintered material (sample labelled TSO\_2) was then isostatically pressed into a rod (7 mm diameter and 60 mm long) and sintered at 1400~$^{\circ}$C in air for several days. The annealed rod was then used for the crystal growth. An analysis (GOF~=~1.51) of the powder X-ray diffraction pattern collected on a ground piece of the feed rod reveals the presence of two phases, C-type Tm$_{2}$Si$_{2}$O$_{7}$ and B-type Tm$_{2}$SiO$_{5}$ (see Fig.~\ref{Fig:Tm2Si2O7_polycrystal_PXRD}(d)). 

The results of our synthesis experiments suggest that the stabilisation of the disilicate phase in the case of thulium is more difficult than for the two other rare-earth disilicate compounds investigated in this study. This is most likely due to the very similar temperature stability ranges of Tm$_{2}$SiO$_{5}$ and Tm$_{2}$Si$_{2}$O$_{7}$ phases~\cite{1970_Felsche,1973_Felsche}.

\subsubsection{Crystal growth}

We have first grown a Tm$_{2}$Si$_{2}$O$_{7}$ crystal boule, in static air atmosphere, at ambient pressure, using growth rates in the range 5-7~mm/h. The feed and seed rods were counter-rotated, each at a rate of 15~rpm. A polycrystalline rod was used as a seed rod. No deposition was observed on the quartz tube surrounding the feed and seed rods, suggesting that Tm$_{2}$Si$_{2}$O$_{7}$ melts congruently. The crystal did not develop any facets and the crystal boule was very fragile. To minimise the thermal stress on the crystal boule and reduce the cracks, the rotation of the seed rod was reduced to 5~rpm. Analysis by X-ray Laue diffraction of the crystalline quality of the Tm$_{2}$Si$_{2}$O$_{7}$ crystal boule obtained revealed the presence of multiple grains.

\begin{figure}[H]
\centering
\includegraphics[width=0.75\columnwidth]{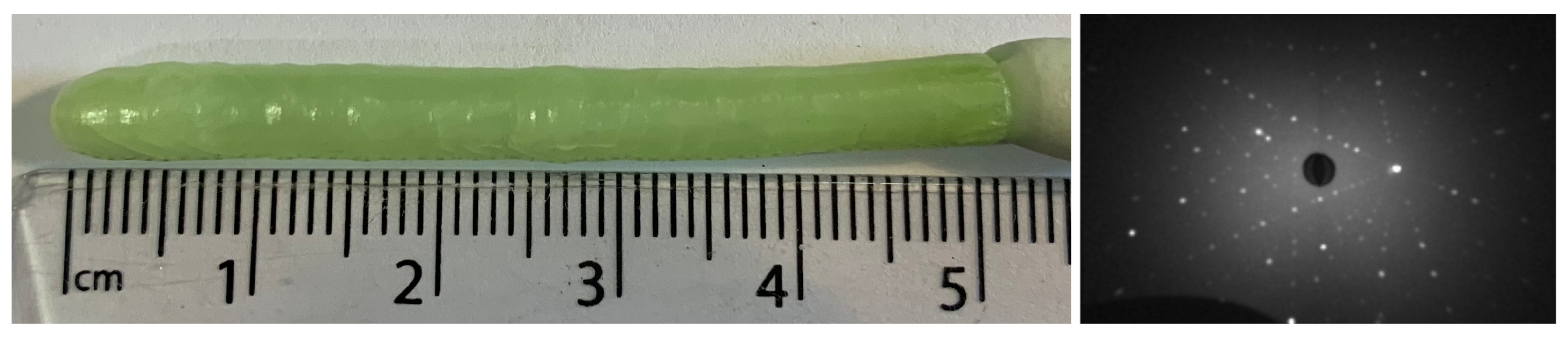}
\caption{Boule of Tm$_{2}$Si$_{2}$O$_{7}$ prepared by the floating zone method in air atmosphere, at ambient pressure, in a flow of air of 1-2~L/min, using a growth speed of 10~mm/h. Also shown is the X-ray Laue pattern of one of the sides of the Tm$_{2}$Si$_{2}$O$_{7}$ crystal boule. }
\label{Fig:Tm2Si2O7_crystal}
\end{figure}

A second crystal growth was performed, using as feed the crystal boule obtained previously. The growth was carried out in air atmosphere, at ambient pressure, in a flow of air of 1-2~L/min, using a growth rate of 10~mm/h (see Fig.~\ref{Fig:Tm2Si2O7_crystal}). The feed and seed rods were counter-rotated, each at a rate of 15-25~rpm. The Tm$_{2}$Si$_{2}$O$_{7}$ crystal boule obtained was $\sim$ 5~mm in diameter and 50~mm long. Thermally generated cracks were observed on the surface of the crystal boule. One facet developed midway during the growth and extended over the remaining length of the grown crystal boule. The thulium disilicate crystal boule obtained was a pale green colour. A visual inspection of the cross-section of the crystal revealed that the boule consists of a semi-transparent shell and an opaque core. Analysis of the X-ray Laue diffraction patterns collected along the sides of this boule of Tm$_{2}$Si$_{2}$O$_{7}$ shows the presence of several grains. Nonetheless, long needle-like single crystals, $\sim$ 10-15$\times$2$\times$1-2~mm$^{3}$, could be isolated from the outermost layer of the crystal boule. These crystals are suitable for magnetic properties measurements of this material.

\begin{figure}[H]
\centering
\includegraphics[width=0.75\columnwidth]{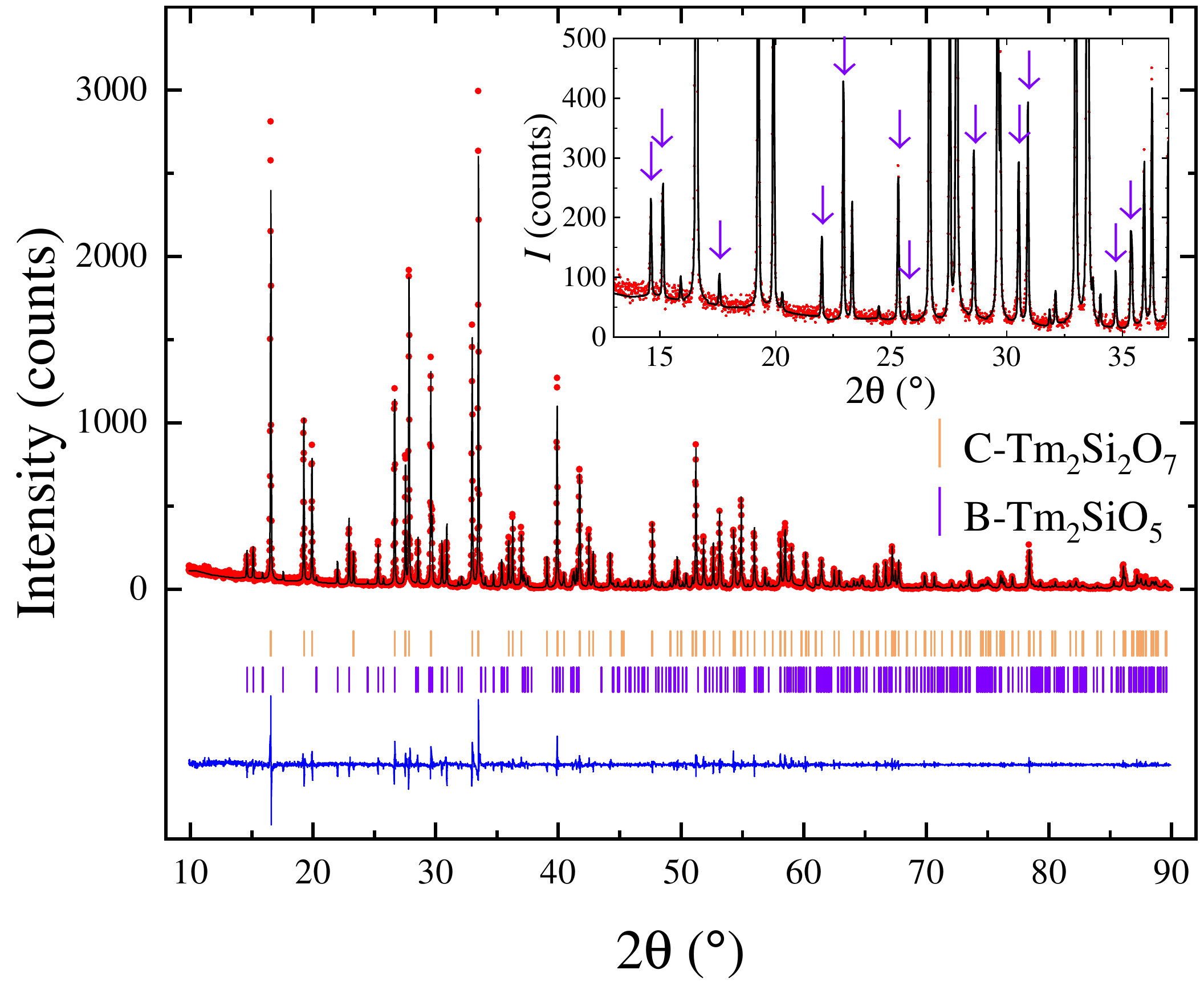}
\caption{Room temperature powder X-ray diffraction pattern of a ground Tm$_{2}$Si$_{2}$O$_{7}$ crystal piece. The experimental profile (red closed circles) and a full profile matching refinement (black solid line) made using the C-type ($C$2/$m$) monoclinic structure are shown, with the difference given by the blue solid line. The reflections of the C-type Tm$_{2}$Si$_{2}$O$_{7}$ structure are indicated by orange "\ding{120}"; purple "\ding{120}" show the reflections belonging to a B-type Tm$_{2}$SiO$_{5}$ (monoclinic $C$2/$c$ structure) impurity. The inset shows the X-ray pattern in a reduced range of the scattering angle 2$\theta$, with the impurity reflections marked by coloured arrows.}
\label{Fig:Tm2Si2O7_crystal_PXRD}
\end{figure}

Phase composition analysis by powder X-ray diffraction (GOF~=~1.50) of a ground cross-section of Tm$_{2}$Si$_{2}$O$_{7}$ boule (made up of both the outer layer and the inner core) reveals that, although the main phase is C-type Tm$_{2}$Si$_{2}$O$_{7}$, there is a small amount of B-type Tm$_{2}$SiO$_{5}$ impurity present in the crystal (see Fig.~\ref{Fig:Tm2Si2O7_crystal_PXRD}). The lattice parameters of monoclinic ($C$2/$m$) thulium disilicate were determined to be $a$~=~6.8276(2)~\AA, $b$~=~8.9105(3)~\AA\ and $c$~=~4.7067(2)~\AA, with the angle $\beta$~=~101.834(2)$^{\circ}$. These values are close to the previously published results on flux grown crystals of C-type Tm$_{2}$Si$_{2}$O$_{7}$~\cite{1970_Felsche,2014_Kahlenberg}.

Composition analysis by EDAX was performed on a cleaved piece from the Tm$_{2}$Si$_{2}$O$_{7}$ crystal boule. The average atomic percentages of Tm, Si and O were 17.1(2)$\%$, 16.7(2)$\%$ and 66.2(3)$\%$ respectively. Given the limitations of this technique, the results are in reasonable agreement with the expected cationic ratio average of 1:1 for Tm:Si for the Tm$_{2}$Si$_{2}$O$_{7}$ phase.

The problems encountered in the preparation of Tm$_{2}$Si$_{2}$O$_{7}$ samples, both in powder and crystal form, are most likely due to the similar ranges of thermal stability of the thulium monoslicate and disilicate, as well as, to the similar melting temperatures of these compounds (see, for example, the Y-Si-O system~\cite{2018_TAbdulJabbar}). Polycrystalline and single crystal synthesis experiments are in progress in order to optimise the conditions and obtain phase C-type Tm$_{2}$Si$_{2}$O$_{7}$ samples.

\subsubsection{Heat capacity}

\begin{figure}[H]
\centering
\includegraphics[width=0.75\columnwidth]{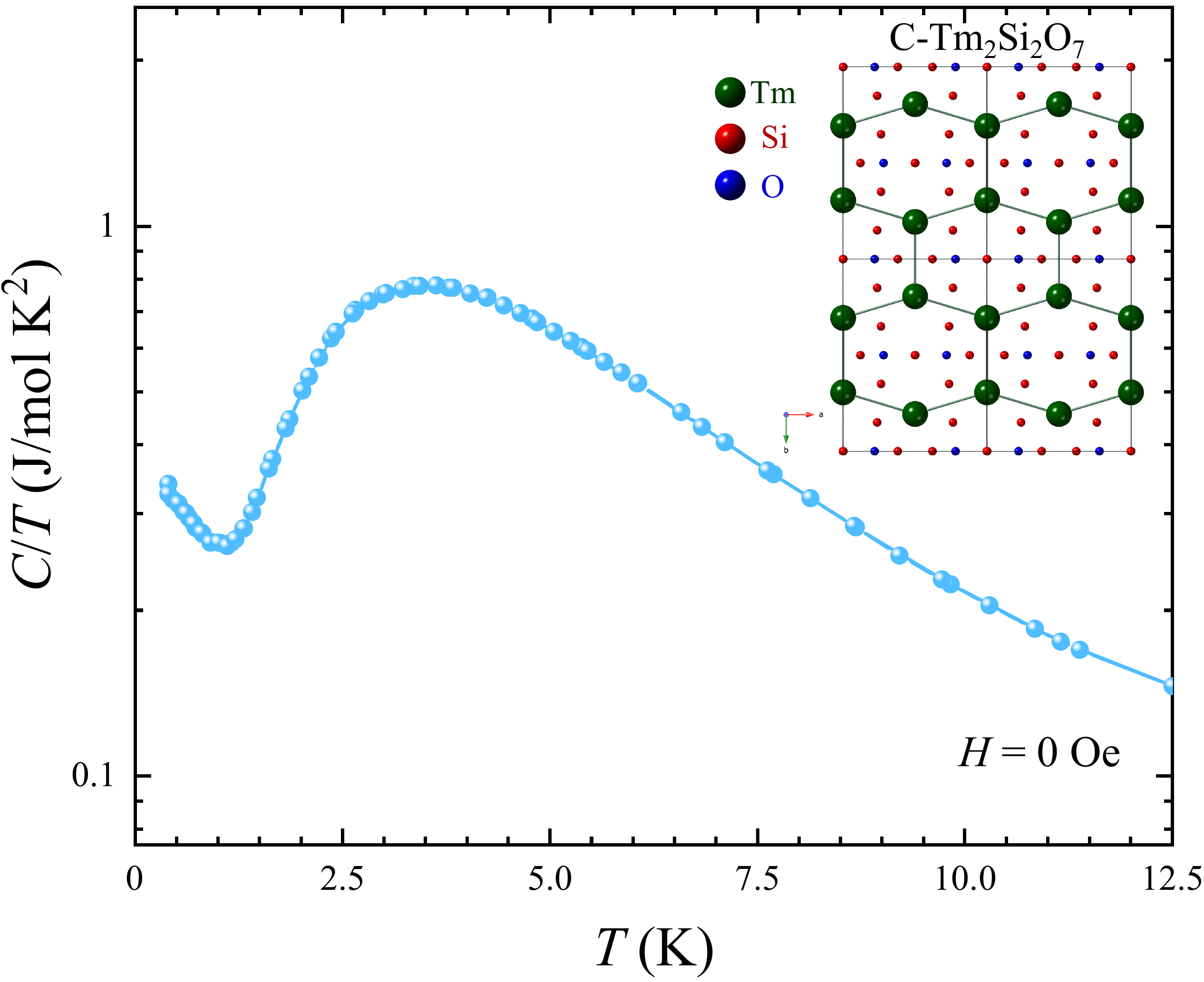}
\caption{$C(T)/T$ for a C-type Tm$_{2}$Si$_{2}$O$_{7}$ crystal piece in zero applied magnetic field. The inset shows the arrangement of the Tm$^{3+}$ magnetic ions in the C-type crystallographic structure, emphasising the presence of distorted honeycombs.}
\label{Fig:Tm2Si2O7_C/TvsT}
\end{figure}

To date, the magnetic properties of C-type Tm$_{2}$Si$_{2}$O$_{7}$ have not been studied. Figure~\ref{Fig:Tm2Si2O7_C/TvsT} shows the heat capacity measurement performed in zero applied magnetic field on an unaligned fragment of a thulium disilicate crystal. Similar to what is observed for C-type Yb$_{2}$Si$_{2}$O$_{7}$, there is no sign of magnetic ordering of Tm$_{2}$Si$_{2}$O$_{7}$ down to $\sim$ 0.4~K. The broad peak centred around 3.5~K corresponds, in all likelihood, to a Schottky anomaly. The increase of $C(T)/T$ below 1~K  suggests the development of short-range magnetic correlations. Detailed investigations at low temperature are being carried out in order to establish the magnetic structure of C-type Tm$_{2}$Si$_{2}$O$_{7}$.

\section{Summary and Conclusions}

Samples of $R_{2}$Si$_{2}$O$_{7}$ (with \textit{R}~=~Er, Ho, and Tm) compounds were first prepared in polycrystalline form by the conventional solid state synthesis method. The analysis by powder X-ray diffraction of the polycrystalline samples prepared revealed the presence of impurities (in the case of holmium and thulium), in the form of $R_{2}$SiO$_{5}$. We have shown that to reduce the formation of the impurity monosilicate phases in the synthesis, it is necessary to start with an excess of SiO$_{2}$. Our synthesis efforts with excess SiO$_{2}$ succeeded in producing samples with considerably reduced levels of the impurity phases and we were able to obtain mainly the desired $R_{2}$Si$_{2}$O$_{7}$ phase. Detailed studies are now being carried out to further optimise the synthesis conditions in order to obtain phase pure powder samples of the rare-earth disilicate compounds, minimising the amount of unreacted SiO$_{2}$. The conditions used for the synthesis of the polycrystalline materials, as well as the results of the phase composition analysis of each sample, are summarised in Table~\ref{Tab:R2Si2O7_polycrystal}.

We have been successful in growing crystals of $R_{2}$Si$_{2}$O$_{7}$ (with \textit{R}~=~Er, Ho, and Tm) using the FZ method. All the rare-earth disilicate compounds appear to melt congruently and no evaporation was observed for the growths. The crystal growths were performed using various growth rates, in air atmosphere. All the rare-earth disilicate crystals are very fragile and tend to have thermally generated cracks. In general, crystal boules of better crystalline quality were obtained using average or high speed of growth. A summary of the crystal growth conditions used for the growth is given in Table~\ref{Tab:R2Si2O7_crystal}.

The quality and composition of the as-grown $R_{2}$Si$_{2}$O$_{7}$ boules were investigated using X-ray diffraction techniques. The lattice parameters determined by powder X-ray diffraction are collected in Table~\ref{Tab:R2Si2O7_param}. Erbium and holmium disilicate crystals were single phase, whereas the thulium crystal boule consists of two chemical phases (monosilicate and disilicate), due to the overlapping ranges of thermal stability of the two thulium silicate compounds, Tm$_{2}$SiO$_{5}$ and Tm$_{2}$Si$_{2}$O$_{7}$. The lattice parameters of the $R_{2}$Si$_{2}$O$_{7}$ crystal boules are in agreement with the previously published results. Despite the difficulties encountered in the preparation of polycrystalline and crystal samples, good size grains could be isolated from the crystal boules to be used for further physical properties characterisation measurements. Further investigations to better understand the stabilisation of $R_{2}$Si$_{2}$O$_{7}$ phases in their crystal form are in progress.

Magnetic properties measurements confirm an antiferromagnetic ordering of Er$^{3+}$ ions at $T_\mathrm{{N}}$~=~1.80(2)~K in D-type Er$_{2}$Si$_{2}$O$_{7}$. We report, for the first time, that E-type Ho$_{2}$Si$_{2}$O$_{7}$ orders antiferromagnetically below $T_\mathrm{{N}}$~=~2.30(5)~K. We also show that Tm$_{2}$Si$_{2}$O$_{7}$ exhibits no signs of long-range magnetic ordering down to $\sim$ 0.4~K, however, short-range magnetic correlations develop below 1~K. Detailed magnetic properties measurements are now being carried out on the $R_{2}$Si$_{2}$O$_{7}$ (with \textit{R}~=~Er, Ho, and Tm) crystals to determine the magnetic ground state of these materials.

\begin{table}[H]
\caption{Summary of the conditions used for the preparation of the $R_{2}$Si$_{2}$O$_{7}$ (with \textit{R}~=~Er, Ho, and Tm) polycrystalline samples. All the samples were sintered for a duration of 48 to 96 hours per step. The results of the phase composition analysis by powder X-ray diffraction are given for each sample.}
\small 
\centering
\begin{tabular}{l l l l l p{5cm}}
\cline{1-6}
\textbf{Chemical}&\textbf{Sample}&\textbf{Sintering}&\textbf{Synthesis}&\textbf{SiO$\mathbf{_{2}}$}&\multirow{2}{*}{\textbf{Phase composition analysis}}\\
\textbf{composition}&\textbf{label}&\textbf{temperature}&\textbf{steps}&\textbf{excess}&\\
&&($^{\circ}$C)&\textbf{number}&(\%)&\\
\cline{1-6}
Er$_{2}$Si$_{2}$O$_{7}$&ESO&1400-1500&4&0&mainly monoclinic D-type Er$_{2}$Si$_{2}$O$_{7}$ and a few peaks of monoclinic B-type Er$_{2}$SiO$_{5}$\\
\cline{1-6}
Ho$_{2}$Si$_{2}$O$_{7}$&HSO\_1&1300&4&0&mixture of monoclinic D-type Ho$_{2}$Si$_{2}$O$_{7}$, monoclinic A-type Ho$_{2}$SiO$_{5}$ and monoclinic B-type Ho$_{2}$SiO$_{5}$\\
\cline{2-6}
&HSO\_1&1300&5&15&mainly monoclinic D-type Ho$_{2}$Si$_{2}$O$_{7}$, a few peaks of monoclinic B-type Ho$_{2}$SiO$_{5}$ and $\alpha$-cristobalite SiO$_{2}$\\
\cline{2-6}
&HSO\_1&1300&6&15&mainly monoclinic D-type Ho$_{2}$Si$_{2}$O$_{7}$, one strong peak of $\alpha$-cristobalite SiO$_{2}$, one peak of monoclinic B-type Ho$_{2}$SiO$_{5}$ and one peak belonging to an unidentified impurity phase\\
\cline{2-6}
&HSO\_2&1400&2&13&mainly monoclinic D-type Ho$_{2}$Si$_{2}$O$_{7}$ and one small peak of $\alpha$-cristobalite SiO$_{2}$\\
\cline{2-6}
&HSO\_3&1400&2&14&mainly monoclinic D-type Ho$_{2}$Si$_{2}$O$_{7}$ and one peak of $\alpha$-cristobalite SiO$_{2}$\\
\cline{2-6}
&HSO\_4&1400&2&15&mainly monoclinic D-type Ho$_{2}$Si$_{2}$O$_{7}$ and one peak of $\alpha$-cristobalite SiO$_{2}$\\
\cline{1-6}
Tm$_{2}$Si$_{2}$O$_{7}$&TSO\_1&1400&5&0&mixture of monoclinic C-type Tm$_{2}$Si$_{2}$O$_{7}$, monoclinic A-type Tm$_{2}$SiO$_{5}$ and monoclinic B-type Tm$_{2}$SiO$_{5}$\\
\cline{2-6}
&TSO\_1&1400&6&17&mainly monoclinic C-type Tm$_{2}$Si$_{2}$O$_{7}$, a few strong peaks of monoclinic monoclinic B-type Tm$_{2}$SiO$_{5}$, and a few peaks of A-type Tm$_{2}$SiO$_{5}$ and $\alpha$-cristobalite SiO$_{2}$\\
\cline{2-6}
&TSO\_1&1400&7&17&mainly monoclinic C-type Tm$_{2}$Si$_{2}$O$_{7}$ and a few peaks of monoclinic monoclinic B-type Tm$_{2}$SiO$_{5}$ and $\alpha$-cristobalite SiO$_{2}$\\
\cline{2-6}
&TSO\_2&1400&4&17&mainly monoclinic C-type Tm$_{2}$Si$_{2}$O$_{7}$ and a few peaks of monoclinic monoclinic B-type Tm$_{2}$SiO$_{5}$\\
\cline{1-6}
\end{tabular}
\label{Tab:R2Si2O7_polycrystal}
\end{table}

\begin{table}[H]
\caption{Summary of the conditions used for the growth of $R_{2}$Si$_{2}$O$_{7}$ (with \textit{R}~=~Er, Ho, and Tm) crystal boules. The optimal growth conditions that allowed us to obtain better quality boules and to isolate good size single crystals fragments for characterisation measurements are marked using \ding{72}.}
\small 
\centering
\begin{tabular}{ccccc}
\cline{1-5}
\multirow{3}{*}{\textbf{$\mathbf{R_{2}}$Si$\mathbf{_{2}}$O$\mathbf{_{7}}$}}&\textbf{Growth}&\textbf{Gas atmosphere/}&\textbf{Feed \& seed}&\multirow{3}{*}{\textbf{Remarks}} \\
 &\textbf{rate}&\textbf{pressure/flow}&\textbf{rotation rate}&  \\
 &(mm/h)&&(rpm)& \\
\cline{1-5}
\multirow{2}{*}{Er$_{2}$Si$_{2}$O$_{7}$}&5-12&air, ambient&10-25&cloudy pink boules\\
	&10-12&air, ambient&20-25&grains $\sim$ 5$\times$5$\times$3~mm$^{3}$ \ding{72}\\
\cline{1-5}
\multirow{2}{*}{Ho$_{2}$Si$_{2}$O$_{7}$}&5-15&air, ambient&15-25&pale orange boules\\
	&8&air, 1-2~bars, 0.1-0.2~L/min&10-25&grains $\sim$ 20$\times$2$\times$1-2~mm$^{3}$ \ding{72}\\
\cline{1-5}
\multirow{2}{*}{Tm$_{2}$Si$_{2}$O$_{7}$}&5-7&air, ambient&15-5&pale green boules\\
	&10&air, ambient, 1-2~L/min&15-25&grains $\sim$ 10-15$\times$2$\times$1-2~mm$^{3}$ \ding{72}\\
\cline{1-5}
\end{tabular}
\label{Tab:R2Si2O7_crystal}
\end{table}

\begin{table}[H]
\caption{Lattice parameters for $R_{2}$Si$_{2}$O$_{7}$ (with \textit{R}~=~Er, Ho, and Tm), refined from the room temperature powder X-ray diffraction data collected on ground pieces isolated from the crystal boules.}
\small 
\centering
\begin{tabular}{ccccccc}
\cline{1-7}
\textbf{Chemical}&\textbf{Structure}&\textbf{Space}&\multicolumn{4}{c}{\textbf{Lattice parameters}}\\
\textbf{composition}&\textbf{(type)}&\textbf{group}&$a$&$b$&$c$&Angle\\
& & &(\AA)&(\AA)&(\AA)&($^{\circ}$)\\
\cline{1-7}
Er$_{2}$Si$_{2}$O$_{7}$&monoclinic (D)&$P$2$_{1}$/$b$&4.6908(2)&5.5615(2)&10.7991(2)&$\gamma$~=~96.040(2)\\
Ho$_{2}$Si$_{2}$O$_{7}$&orthorhombic (E)&$Pna2_{1}$&13.6770(3)&5.0235(3)&8.1598(3)&-\\
Tm$_{2}$Si$_{2}$O$_{7}$&monoclinic (C)&$C$2/$m$&6.8276(2)&8.9105(3)&4.7067(2)&$\beta$~=~101.834(2)\\
\cline{1-7}
\end{tabular}
\label{Tab:R2Si2O7_param}
\end{table}

\begin{acknowledgement}

Financial support was provided by EPSRC, UK through Grant EP/T005963/1. The authors would like to acknowledge the contributions of S. Donner, J. Mileson, M. Minney, and G. Palmer to the preparation of rare-earth silicate compounds through their involvement with undergraduate projects. The authors would also like to thank S. J. York for the EDAX compositional analysis.

\end{acknowledgement}

\bibliography{R2Si2O7_paper}

\providecommand{\latin}[1]{#1}
\makeatletter
\providecommand{\doi}
  {\begingroup\let\do\@makeother\dospecials
  \catcode`\{=1 \catcode`\}=2 \doi@aux}
\providecommand{\doi@aux}[1]{\endgroup\texttt{#1}}
\makeatother
\providecommand*\mcitethebibliography{\thebibliography}
\csname @ifundefined\endcsname{endmcitethebibliography}
  {\let\endmcitethebibliography\endthebibliography}{}
\begin{mcitethebibliography}{47}
\providecommand*\natexlab[1]{#1}
\providecommand*\mciteSetBstSublistMode[1]{}
\providecommand*\mciteSetBstMaxWidthForm[2]{}
\providecommand*\mciteBstWouldAddEndPuncttrue
  {\def\EndOfBibitem{\unskip.}}
\providecommand*\mciteBstWouldAddEndPunctfalse
  {\let\EndOfBibitem\relax}
\providecommand*\mciteSetBstMidEndSepPunct[3]{}
\providecommand*\mciteSetBstSublistLabelBeginEnd[3]{}
\providecommand*\EndOfBibitem{}
\mciteSetBstSublistMode{f}
\mciteSetBstMaxWidthForm{subitem}{(\alph{mcitesubitemcount})}
\mciteSetBstSublistLabelBeginEnd
  {\mcitemaxwidthsubitemform\space}
  {\relax}
  {\relax}

\bibitem[Ito and Johnson(1968)Ito, and Johnson]{1968_Ito}
Ito,~J.; Johnson,~H. Synthesis and study of yttrialite. \emph{Am. Mineral.}
  \textbf{1968}, \emph{53}, 1940--1952\relax
\mciteBstWouldAddEndPuncttrue
\mciteSetBstMidEndSepPunct{\mcitedefaultmidpunct}
{\mcitedefaultendpunct}{\mcitedefaultseppunct}\relax
\EndOfBibitem
\bibitem[Felsche(1970)]{1970_Felsche}
Felsche,~J. Polymorphism and crystal data of the rare-earth disilicates of type
  R.E.$_{2}$Si$_{2}$O$_{7}$. \emph{J. Less-Common Met.} \textbf{1970},
  \emph{21}, 1--14\relax
\mciteBstWouldAddEndPuncttrue
\mciteSetBstMidEndSepPunct{\mcitedefaultmidpunct}
{\mcitedefaultendpunct}{\mcitedefaultseppunct}\relax
\EndOfBibitem
\bibitem[Felsche(1973)]{1973_Felsche}
Felsche,~J. The crystal chemistry of the rare-earth silicates. \emph{Struct.
  Bonding (Berlin, Ger.)} \textbf{1973}, \emph{13}, 99--197\relax
\mciteBstWouldAddEndPuncttrue
\mciteSetBstMidEndSepPunct{\mcitedefaultmidpunct}
{\mcitedefaultendpunct}{\mcitedefaultseppunct}\relax
\EndOfBibitem
\bibitem[Liu and Fleet(2002)Liu, and Fleet]{2002_Liu}
Liu,~X.; Fleet,~M.~E. High-pressure synthesis of a La orthosilicate and Nd, Gd,
  and Dy disilicates. \emph{J. Phys.: Condens. Matter} \textbf{2002},
  \emph{14}, 11223--11226\relax
\mciteBstWouldAddEndPuncttrue
\mciteSetBstMidEndSepPunct{\mcitedefaultmidpunct}
{\mcitedefaultendpunct}{\mcitedefaultseppunct}\relax
\EndOfBibitem
\bibitem[Bretheau-Raynal \latin{et~al.}(1980)Bretheau-Raynal, Tercier, Blanzat,
  and Drifford]{1980_Bretheau-Raynal}
Bretheau-Raynal,~F.; Tercier,~N.; Blanzat,~B.; Drifford,~M. Synthesis and
  spectroscopic study of lutetium pyrosilicate single crystals doped with
  trivalent europium. \emph{Mater. Res. Bull.} \textbf{1980}, \emph{15},
  639--646\relax
\mciteBstWouldAddEndPuncttrue
\mciteSetBstMidEndSepPunct{\mcitedefaultmidpunct}
{\mcitedefaultendpunct}{\mcitedefaultseppunct}\relax
\EndOfBibitem
\bibitem[Pauwels \latin{et~al.}(2000)Pauwels, Le~Masson, Viana, Kahn-Harari,
  van Loef, Dorenbos, and van Eijk]{2000_Pauwels}
Pauwels,~D.; Le~Masson,~N.; Viana,~B.; Kahn-Harari,~A.; van Loef,~E.;
  Dorenbos,~P.; van Eijk,~C. A novel inorganic scintillator:
  Lu$_{2}$Si$_{2}$O$_{7}$:Ce$^{3+}$ (LPS). \emph{IEEE Trans. Nucl. Sci.}
  \textbf{2000}, \emph{47}, 1787--1790\relax
\mciteBstWouldAddEndPuncttrue
\mciteSetBstMidEndSepPunct{\mcitedefaultmidpunct}
{\mcitedefaultendpunct}{\mcitedefaultseppunct}\relax
\EndOfBibitem
\bibitem[Feng \latin{et~al.}(2010)Feng, Ding, Li, Lu, Pan, Chen, and
  Ren]{2010_Feng}
Feng,~H.; Ding,~D.; Li,~H.; Lu,~S.; Pan,~S.; Chen,~X.; Ren,~G. Growth and
  luminescence characteristics of cerium-doped yttrium pyrosilicate single
  crystal. \emph{J. Alloys Compd.} \textbf{2010}, \emph{489}, 645--649\relax
\mciteBstWouldAddEndPuncttrue
\mciteSetBstMidEndSepPunct{\mcitedefaultmidpunct}
{\mcitedefaultendpunct}{\mcitedefaultseppunct}\relax
\EndOfBibitem
\bibitem[He \latin{et~al.}(2012)He, Guohao, Yuntao, Jun, Qiuhong, Jianjun,
  Mitch, and Chenlong]{2012_He}
He,~F.; Guohao,~R.; Yuntao,~W.; Jun,~X.; Qiuhong,~Y.; Jianjun,~X.; Mitch,~C.;
  Chenlong,~C. Optical and thermoluminescence properties of
  Lu$_{2}$Si$_{2}$O$_{7}$:Pr single crystal. \emph{J. Rare Earths}
  \textbf{2012}, \emph{30}, 775--779\relax
\mciteBstWouldAddEndPuncttrue
\mciteSetBstMidEndSepPunct{\mcitedefaultmidpunct}
{\mcitedefaultendpunct}{\mcitedefaultseppunct}\relax
\EndOfBibitem
\bibitem[Ferna\'{n}dez-Carri\'{o}n
  \latin{et~al.}(2013)Ferna\'{n}dez-Carri\'{o}n, Allix, Oca\~{n}a,
  Garc\'{i}a-Sevillano, Cusso, Fitch, Suard, and
  Becerro]{2013_Fernandez-Carrion}
Ferna\'{n}dez-Carri\'{o}n,~A.~J.; Allix,~M.; Oca\~{n}a,~M.;
  Garc\'{i}a-Sevillano,~J.; Cusso,~F.; Fitch,~A.~N.; Suard,~E.; Becerro,~A.~I.
  Crystal Structures and Photoluminescence across the
  La$_{2}$Si$_{2}$O$_{7}$-Ho$_{2}$Si$_{2}$O$_{7}$ System. \emph{Inorg. Chem.}
  \textbf{2013}, \emph{52}, 13469--13479\relax
\mciteBstWouldAddEndPuncttrue
\mciteSetBstMidEndSepPunct{\mcitedefaultmidpunct}
{\mcitedefaultendpunct}{\mcitedefaultseppunct}\relax
\EndOfBibitem
\bibitem[Lee \latin{et~al.}(2005)Lee, Fox, and Bansal]{2005_Lee}
Lee,~K.~N.; Fox,~D.~S.; Bansal,~N.~P. Rare earth silicate environmental barrier
  coatings for SiC/SiC composites and Si$_{3}$N$_{4}$ ceramics. \emph{J. Eur.
  Ceram. Soc.} \textbf{2005}, \emph{25}, 1705--1715\relax
\mciteBstWouldAddEndPuncttrue
\mciteSetBstMidEndSepPunct{\mcitedefaultmidpunct}
{\mcitedefaultendpunct}{\mcitedefaultseppunct}\relax
\EndOfBibitem
\bibitem[Sun \latin{et~al.}(2008)Sun, Zhou, Wang, and Li]{2008_Sun}
Sun,~Z.; Zhou,~Y.; Wang,~J.; Li,~M. Thermal Properties and Thermal Shock
  Resistance of $\gamma$ - Y$_{2}$Si$_{2}$O$_{7}$. \emph{J. Am. Ceram. Soc.}
  \textbf{2008}, \emph{91}, 2623--2629\relax
\mciteBstWouldAddEndPuncttrue
\mciteSetBstMidEndSepPunct{\mcitedefaultmidpunct}
{\mcitedefaultendpunct}{\mcitedefaultseppunct}\relax
\EndOfBibitem
\bibitem[Sun \latin{et~al.}(2009)Sun, Li, and Zhou]{2009_Sun}
Sun,~Z.; Li,~M.; Zhou,~Y. Thermal properties of single-phase Y$_{2}$SiO$_{5}$.
  \emph{J. Eur. Ceram. Soc.} \textbf{2009}, \emph{29}, 551--557\relax
\mciteBstWouldAddEndPuncttrue
\mciteSetBstMidEndSepPunct{\mcitedefaultmidpunct}
{\mcitedefaultendpunct}{\mcitedefaultseppunct}\relax
\EndOfBibitem
\bibitem[Zhou \latin{et~al.}(2013)Zhou, Zhao, Wang, Sun, Zheng, and
  Wang]{2013_Zhou}
Zhou,~Y.; Zhao,~C.; Wang,~F.; Sun,~Y.; Zheng,~L.; Wang,~X. Theoretical
  Prediction and Experimental Investigation on the Thermal and Mechanical
  Properties of Bulk $\beta$ - Yb$_{2}$Si$_{2}$O$_{7}$. \emph{J. Am. Ceram.
  Soc.} \textbf{2013}, \emph{96}, 3891--3900\relax
\mciteBstWouldAddEndPuncttrue
\mciteSetBstMidEndSepPunct{\mcitedefaultmidpunct}
{\mcitedefaultendpunct}{\mcitedefaultseppunct}\relax
\EndOfBibitem
\bibitem[Tian \latin{et~al.}(2016)Tian, Zheng, Li, Li, and Wang]{2016_Tian}
Tian,~Z.; Zheng,~L.; Li,~Z.; Li,~J.; Wang,~J. Exploration of the low thermal
  conductivities of $\gamma$ - Y$_{2}$Si$_{2}$O$_{7}$, $\beta$ -
  Y$_{2}$Si$_{2}$O$_{7}$, $\beta$ - Yb$_{2}$Si$_{2}$O$_{7}$, and $\beta$ -
  Lu$_{2}$Si$_{2}$O$_{7}$ as novel environmental barrier coating candidates.
  \emph{J. Eur. Ceram. Soc.} \textbf{2016}, \emph{36}, 2813--2823\relax
\mciteBstWouldAddEndPuncttrue
\mciteSetBstMidEndSepPunct{\mcitedefaultmidpunct}
{\mcitedefaultendpunct}{\mcitedefaultseppunct}\relax
\EndOfBibitem
\bibitem[Luo \latin{et~al.}(2018)Luo, Sun, Wang, Wu, Lv, and Wang]{2018_Luo}
Luo,~Y.; Sun,~L.; Wang,~W.; Wu,~Z.; Lv,~X.; Wang,~J. Material-genome
  perspective towards tunable thermal expansion of rare-earth di-silicates.
  \emph{J. Eur. Ceram. Soc.} \textbf{2018}, \emph{38}, 3547--3554\relax
\mciteBstWouldAddEndPuncttrue
\mciteSetBstMidEndSepPunct{\mcitedefaultmidpunct}
{\mcitedefaultendpunct}{\mcitedefaultseppunct}\relax
\EndOfBibitem
\bibitem[Nair \latin{et~al.}(2019)Nair, DeLazzer, Reeder, Sikorski, Hester, and
  Ross]{2019_Nair}
Nair,~H.~S.; DeLazzer,~T.; Reeder,~T.; Sikorski,~A.; Hester,~G.; Ross,~K.~A.
  Crystal Growth of Quantum Magnets in the Rare-Earth Pyrosilicate Family
  $R_{2}$Si$_{2}$O$_{7}$ ($R$~=~Yb, Er) Using the Optical Floating Zone Method.
  \emph{Crystals} \textbf{2019}, \emph{9}, 196\relax
\mciteBstWouldAddEndPuncttrue
\mciteSetBstMidEndSepPunct{\mcitedefaultmidpunct}
{\mcitedefaultendpunct}{\mcitedefaultseppunct}\relax
\EndOfBibitem
\bibitem[Hester \latin{et~al.}(2019)Hester, Nair, Reeder, Yahne, DeLazzer,
  Berges, Ziat, Neilson, Aczel, Sala, Quilliam, and Ross]{2019_Hester}
Hester,~G.; Nair,~H.; Reeder,~T.; Yahne,~D.; DeLazzer,~T.; Berges,~L.;
  Ziat,~D.; Neilson,~J.; Aczel,~A.; Sala,~G.; Quilliam,~J.; Ross,~K. Novel
  Strongly Spin-Orbit Coupled Quantum Dimer Magnet: Yb$_{2}$Si$_{2}$O$_{7}$.
  \emph{Phys. Rev. Lett.} \textbf{2019}, \emph{123}, 027201\relax
\mciteBstWouldAddEndPuncttrue
\mciteSetBstMidEndSepPunct{\mcitedefaultmidpunct}
{\mcitedefaultendpunct}{\mcitedefaultseppunct}\relax
\EndOfBibitem
\bibitem[Flynn \latin{et~al.}(2020)Flynn, Baker, Jindal, and Singh]{2020_Flynn}
Flynn,~M.~O.; Baker,~T.~E.; Jindal,~S.; Singh,~R. R.~P. On two phases inside
  the Bose condensation dome of Yb$_{2}$Si$_{2}$O$_{7}$. \emph{arXiv.org,
  e-Print Arch., Condens. Matter} \textbf{2020}, arXiv:2001.08219\relax
\mciteBstWouldAddEndPuncttrue
\mciteSetBstMidEndSepPunct{\mcitedefaultmidpunct}
{\mcitedefaultendpunct}{\mcitedefaultseppunct}\relax
\EndOfBibitem
\bibitem[Smolin and Shepelev(1970)Smolin, and Shepelev]{1970_Smolin}
Smolin,~Y.~I.; Shepelev,~Y.~F. The crystal structures of the rare earth
  pyrosilicates. \emph{Acta Crystallogr., Sect. B: Struct. Sci., Cryst. Eng.
  Mater.} \textbf{1970}, \emph{26}, 484--492\relax
\mciteBstWouldAddEndPuncttrue
\mciteSetBstMidEndSepPunct{\mcitedefaultmidpunct}
{\mcitedefaultendpunct}{\mcitedefaultseppunct}\relax
\EndOfBibitem
\bibitem[Wanklyn \latin{et~al.}(1974)Wanklyn, Wondre, Ansell, and
  Davison]{1974_Wanklyn}
Wanklyn,~B.~M.; Wondre,~F.~R.; Ansell,~G.~B.; Davison,~W. Flux growth of rare
  earth silicates and aluminosilicates. \emph{J. Mater. Sci.} \textbf{1974},
  \emph{9}, 2007--2014\relax
\mciteBstWouldAddEndPuncttrue
\mciteSetBstMidEndSepPunct{\mcitedefaultmidpunct}
{\mcitedefaultendpunct}{\mcitedefaultseppunct}\relax
\EndOfBibitem
\bibitem[Wanklyn(1978)]{1978_Wanklyn}
Wanklyn,~B.~M. Effects of modifying starting compositions for flux growth.
  \emph{J. Cryst. Growth} \textbf{1978}, \emph{43}, 336--344\relax
\mciteBstWouldAddEndPuncttrue
\mciteSetBstMidEndSepPunct{\mcitedefaultmidpunct}
{\mcitedefaultendpunct}{\mcitedefaultseppunct}\relax
\EndOfBibitem
\bibitem[Maqsood \latin{et~al.}(1979)Maqsood, Wanklyn, and
  Garton]{1979_Maqsood}
Maqsood,~A.; Wanklyn,~B.~M.; Garton,~G. Flux growth of polymorphic rare-earth
  disilicates, $R_{2}$Si$_{2}$O$_{7}$ ($R$~=~Tm, Er, Ho, Dy). \emph{J. Cryst.
  Growth} \textbf{1979}, \emph{46}, 671--680\relax
\mciteBstWouldAddEndPuncttrue
\mciteSetBstMidEndSepPunct{\mcitedefaultmidpunct}
{\mcitedefaultendpunct}{\mcitedefaultseppunct}\relax
\EndOfBibitem
\bibitem[{N{\o}rlund Christensen} \latin{et~al.}(1997){N{\o}rlund Christensen},
  Hazell, and Hewat]{1997_NorlundChristensen}
{N{\o}rlund Christensen},~A.; Hazell,~R.~G.; Hewat,~A.~W. Synthesis, Crystal
  Growth and Structure Investigations of Rare-Earth Disilicates and Rare-Earth
  Oxyapatites. \emph{Acta Chem. Scand.} \textbf{1997}, \emph{51}, 37--43\relax
\mciteBstWouldAddEndPuncttrue
\mciteSetBstMidEndSepPunct{\mcitedefaultmidpunct}
{\mcitedefaultendpunct}{\mcitedefaultseppunct}\relax
\EndOfBibitem
\bibitem[Maqsood(2000)]{2000_Maqsood}
Maqsood,~A. Single crystal growth of polymorphic Er$_{2}$Si$_{2}$O$_{7}$
  ceramics. \emph{J. Mater. Sci. Lett.} \textbf{2000}, \emph{19},
  711--712\relax
\mciteBstWouldAddEndPuncttrue
\mciteSetBstMidEndSepPunct{\mcitedefaultmidpunct}
{\mcitedefaultendpunct}{\mcitedefaultseppunct}\relax
\EndOfBibitem
\bibitem[Kahlenberg and Aichholzer(2014)Kahlenberg, and
  Aichholzer]{2014_Kahlenberg}
Kahlenberg,~V.; Aichholzer,~P. Thortveitite-type Tm$_{2}$Si$_{2}$O$_{7}$.
  \emph{Acta Crystallogr., Sect. E: Struct. Rep. Online} \textbf{2014},
  \emph{70}, i34--i35\relax
\mciteBstWouldAddEndPuncttrue
\mciteSetBstMidEndSepPunct{\mcitedefaultmidpunct}
{\mcitedefaultendpunct}{\mcitedefaultseppunct}\relax
\EndOfBibitem
\bibitem[Chi \latin{et~al.}(1998)Chi, Chen, Zhuang, and Huang]{1998_Chi}
Chi,~L.-S.; Chen,~H.-Y.; Zhuang,~H.-H.; Huang,~J.-S. Synthesis and Crystal
  Structure of Er$_{2}$Si$_{2}$O$_{7}$. \emph{Chin. J. Struct. Chem.}
  \textbf{1998}, \emph{17}, 24--26\relax
\mciteBstWouldAddEndPuncttrue
\mciteSetBstMidEndSepPunct{\mcitedefaultmidpunct}
{\mcitedefaultendpunct}{\mcitedefaultseppunct}\relax
\EndOfBibitem
\bibitem[Horiai \latin{et~al.}(2016)Horiai, Kurosawa, Murakami, Pejchal,
  Yamaji, Shoji, Chani, Ohashi, Kamada, Yokota, and Yoshikawa]{2016_Horiai}
Horiai,~T.; Kurosawa,~S.; Murakami,~R.; Pejchal,~J.; Yamaji,~A.; Shoji,~Y.;
  Chani,~V.; Ohashi,~Y.; Kamada,~K.; Yokota,~Y.; Yoshikawa,~A. Crystal growth
  and luminescence properties of Yb$_{2}$Si$_{2}$O$_{7}$ infra-red emission
  scintillator. \emph{Opt. Mater. (Amsterdam, Neth.)} \textbf{2016}, \emph{58},
  14--17\relax
\mciteBstWouldAddEndPuncttrue
\mciteSetBstMidEndSepPunct{\mcitedefaultmidpunct}
{\mcitedefaultendpunct}{\mcitedefaultseppunct}\relax
\EndOfBibitem
\bibitem[Balakrishnan \latin{et~al.}(1998)Balakrishnan, Petrenko, Lees, and
  Paul]{1998_Balakrishnan}
Balakrishnan,~G.; Petrenko,~O.~A.; Lees,~M.~R.; Paul,~D.~M. Single crystal
  growth of rare earth titanate pyrochlores. \emph{J. Phys.: Condens. Matter}
  \textbf{1998}, \emph{10}, L723--L725\relax
\mciteBstWouldAddEndPuncttrue
\mciteSetBstMidEndSepPunct{\mcitedefaultmidpunct}
{\mcitedefaultendpunct}{\mcitedefaultseppunct}\relax
\EndOfBibitem
\bibitem[Koohpayeh \latin{et~al.}(2008)Koohpayeh, Fort, and
  Abell]{2008_Koohpayeh}
Koohpayeh,~S.~M.; Fort,~D.; Abell,~J.~S. The optical floating zone technique: A
  review of experimental procedures with special reference to oxides.
  \emph{Prog. Cryst. Growth Charact. Mater.} \textbf{2008}, \emph{54},
  121--137\relax
\mciteBstWouldAddEndPuncttrue
\mciteSetBstMidEndSepPunct{\mcitedefaultmidpunct}
{\mcitedefaultendpunct}{\mcitedefaultseppunct}\relax
\EndOfBibitem
\bibitem[Dabkowska and Dabkowski(2015)Dabkowska, and Dabkowski]{2015_Dabkowska}
Dabkowska,~H.; Dabkowski,~A. \emph{Handbook of Crystal Growth (Second
  Edition)}; Elsevier, 2015; pp 281--329\relax
\mciteBstWouldAddEndPuncttrue
\mciteSetBstMidEndSepPunct{\mcitedefaultmidpunct}
{\mcitedefaultendpunct}{\mcitedefaultseppunct}\relax
\EndOfBibitem
\bibitem[Maqsood and ul~Haq(1987)Maqsood, and ul~Haq]{1987_Maqsood}
Maqsood,~A.; ul~Haq,~I. Preparation of rare-earth disilicates and their X-ray
  diffraction studies. \emph{J. Mater. Sci. Lett.} \textbf{1987}, \emph{6},
  1095–1097\relax
\mciteBstWouldAddEndPuncttrue
\mciteSetBstMidEndSepPunct{\mcitedefaultmidpunct}
{\mcitedefaultendpunct}{\mcitedefaultseppunct}\relax
\EndOfBibitem
\bibitem[Maqsood(1997)]{1997_Maqsood}
Maqsood,~A. Phase transformations in Er$_{2}$Si$_{2}$O$_{7}$ ceramics. \emph{J.
  Mater. Sci. Lett.} \textbf{1997}, \emph{16}, 837--840\relax
\mciteBstWouldAddEndPuncttrue
\mciteSetBstMidEndSepPunct{\mcitedefaultmidpunct}
{\mcitedefaultendpunct}{\mcitedefaultseppunct}\relax
\EndOfBibitem
\bibitem[Maqsood(2009)]{2009_Maqsood}
Maqsood,~A. Phase transformations in Ho$_{2}$Si$_{2}$O$_{7}$ ceramics. \emph{J.
  Alloys Compd.} \textbf{2009}, \emph{471}, 432--434\relax
\mciteBstWouldAddEndPuncttrue
\mciteSetBstMidEndSepPunct{\mcitedefaultmidpunct}
{\mcitedefaultendpunct}{\mcitedefaultseppunct}\relax
\EndOfBibitem
\bibitem[{Ciomaga Hatnean} \latin{et~al.}(2016){Ciomaga Hatnean}, Decorse,
  Lees, Petrenko, and Balakrishnan]{2016_CiomagaHatnean}
{Ciomaga Hatnean},~M.; Decorse,~C.; Lees,~M.~R.; Petrenko,~O.~A.;
  Balakrishnan,~G. Zirconate pyrochlore frustrated magnets: crystal growth by
  the floating zone technique. \emph{Crystals} \textbf{2016}, \emph{6},
  79\relax
\mciteBstWouldAddEndPuncttrue
\mciteSetBstMidEndSepPunct{\mcitedefaultmidpunct}
{\mcitedefaultendpunct}{\mcitedefaultseppunct}\relax
\EndOfBibitem
\bibitem[Sibille \latin{et~al.}(2017)Sibille, Lhotel, {Ciomaga Hatnean},
  Nilsen, Ehlers, Cervellino, Ressouche, Frontzek, Zaharko, Pomjakushin, Stuhr,
  Walker, Adroja, Luetkens, Baines, Amato, Balakrishnan, Fennell, and
  Kenzelmann]{2017_Sibille}
Sibille,~R.; Lhotel,~E.; {Ciomaga Hatnean},~M.; Nilsen,~G.~J.; Ehlers,~G.;
  Cervellino,~A.; Ressouche,~E.; Frontzek,~M.; Zaharko,~O.; Pomjakushin,~V.;
  Stuhr,~U.; Walker,~H.~C.; Adroja,~D.~T.; Luetkens,~H.; Baines,~C.; Amato,~A.;
  Balakrishnan,~G.; Fennell,~T.; Kenzelmann,~M. Coulomb spin liquid in
  anion-disordered pyrochlore Tb$_{2}$Hf$_{2}$O$_{7}$. \emph{Nat. Commun.}
  \textbf{2017}, \emph{8}, 892\relax
\mciteBstWouldAddEndPuncttrue
\mciteSetBstMidEndSepPunct{\mcitedefaultmidpunct}
{\mcitedefaultendpunct}{\mcitedefaultseppunct}\relax
\EndOfBibitem
\bibitem[Maqsood(1981)]{1981_Maqsood}
Maqsood,~A. Magnetic properties of D-Er$_{2}$Si$_{2}$O$_{7}$ at low
  temperatures. \emph{J. Mater. Sci.} \textbf{1981}, \emph{16},
  2198--2204\relax
\mciteBstWouldAddEndPuncttrue
\mciteSetBstMidEndSepPunct{\mcitedefaultmidpunct}
{\mcitedefaultendpunct}{\mcitedefaultseppunct}\relax
\EndOfBibitem
\bibitem[Leask \latin{et~al.}(1986)Leask, Tapster, and Wells]{1986_Leask}
Leask,~M. J.~M.; Tapster,~P.~R.; Wells,~M.~R. Magnetic properties of
  D-Er$_{2}$Si$_{2}$O$_{7}$. \emph{J. Phys. C: Solid State Phys.}
  \textbf{1986}, \emph{19}, 1173--1187\relax
\mciteBstWouldAddEndPuncttrue
\mciteSetBstMidEndSepPunct{\mcitedefaultmidpunct}
{\mcitedefaultendpunct}{\mcitedefaultseppunct}\relax
\EndOfBibitem
\bibitem[Rodr\'{i}guez-Carvajal(1993)]{1993_RodriguezCarvajal}
Rodr\'{i}guez-Carvajal,~J. Recent advances in magnetic structure determination
  by neutron powder diffraction. \emph{Phys. B (Amsterdam, Neth.)}
  \textbf{1993}, \emph{192}, 55--69\relax
\mciteBstWouldAddEndPuncttrue
\mciteSetBstMidEndSepPunct{\mcitedefaultmidpunct}
{\mcitedefaultendpunct}{\mcitedefaultseppunct}\relax
\EndOfBibitem
\bibitem[Phanon and \v{C}er\'{n}y(2008)Phanon, and \v{C}er\'{n}y]{2008_Phanon}
Phanon,~D.; \v{C}er\'{n}y,~R. Crystal structure of the B‐type dierbium oxide
  ortho‐oxosilicate Er$_{2}$O[SiO$_{4}$]. \emph{Z. Anorg. Allg. Chem.}
  \textbf{2008}, \emph{634}, 1833--1835\relax
\mciteBstWouldAddEndPuncttrue
\mciteSetBstMidEndSepPunct{\mcitedefaultmidpunct}
{\mcitedefaultendpunct}{\mcitedefaultseppunct}\relax
\EndOfBibitem
\bibitem[Petrenko \latin{et~al.}(2020)Petrenko, {Ciomaga Hatnean}, Manuel,
  Orlandi, Khalyavin, and Balakrishnan]{2020_Petrenko}
Petrenko,~O.~A.; {Ciomaga Hatnean},~M.; Manuel,~P.; Orlandi,~F.;
  Khalyavin,~D.~D.; Balakrishnan,~G. in preparation. \textbf{2020}, \relax
\mciteBstWouldAddEndPunctfalse
\mciteSetBstMidEndSepPunct{\mcitedefaultmidpunct}
{}{\mcitedefaultseppunct}\relax
\EndOfBibitem
\bibitem[Maqsood(1984)]{1984_Maqsood}
Maqsood,~A. Single crystal preparation of the rare earth oxyorthosilicates
  $R_{2}$SiO$_{5}$ ($R$=Er, Ho, Dy) by a flux method. \emph{J. Mater. Sci.
  Lett.} \textbf{1984}, \emph{3}, 65--67\relax
\mciteBstWouldAddEndPuncttrue
\mciteSetBstMidEndSepPunct{\mcitedefaultmidpunct}
{\mcitedefaultendpunct}{\mcitedefaultseppunct}\relax
\EndOfBibitem
\bibitem[Chaklader and Roberts(1961)Chaklader, and Roberts]{1961_Chaklader}
Chaklader,~A. C.~D.; Roberts,~A.~L. Transformation of Quartz to Cristobalite.
  \emph{J. Am. Ceram. Soc.} \textbf{1961}, \emph{44}, 35--41\relax
\mciteBstWouldAddEndPuncttrue
\mciteSetBstMidEndSepPunct{\mcitedefaultmidpunct}
{\mcitedefaultendpunct}{\mcitedefaultseppunct}\relax
\EndOfBibitem
\bibitem[Downs and Palmer(1994)Downs, and Palmer]{1994_Downs}
Downs,~R.~T.; Palmer,~D.~C. The pressure behavior of $\alpha$ cristobalite.
  \emph{Am. Mineral.} \textbf{1994}, \emph{79}, 9--14\relax
\mciteBstWouldAddEndPuncttrue
\mciteSetBstMidEndSepPunct{\mcitedefaultmidpunct}
{\mcitedefaultendpunct}{\mcitedefaultseppunct}\relax
\EndOfBibitem
\bibitem[Wang \latin{et~al.}(2001)Wang, Tian, Li, Liao, and Jing]{2001_Tian}
Wang,~J.; Tian,~S.; Li,~G.; Liao,~F.; Jing,~X. Preparation and X-ray
  characterization of low-temperature phases of R$_{2}$SiO$_{5}$ (R = rare
  earth elements). \emph{Mater. Res. Bull.} \textbf{2001}, \emph{36},
  1855--1861\relax
\mciteBstWouldAddEndPuncttrue
\mciteSetBstMidEndSepPunct{\mcitedefaultmidpunct}
{\mcitedefaultendpunct}{\mcitedefaultseppunct}\relax
\EndOfBibitem
\bibitem[Tian \latin{et~al.}(2016)Tian, Zheng, Wang, Wan, Li, and
  Wang]{2016_Tian2}
Tian,~Z.; Zheng,~L.; Wang,~J.; Wan,~P.; Li,~J.; Wang,~J. Theoretical and
  experimental determination of the major thermo-mechanical properties of
  RE$_{2}$SiO$_{5}$ (RE = Tb, Dy, Ho, Er, Tm, Yb, Lu, and Y) for environmental
  and thermal barrier coating applications. \emph{J. Eur. Ceram. Soc.}
  \textbf{2016}, \emph{36}, 189--202\relax
\mciteBstWouldAddEndPuncttrue
\mciteSetBstMidEndSepPunct{\mcitedefaultmidpunct}
{\mcitedefaultendpunct}{\mcitedefaultseppunct}\relax
\EndOfBibitem
\bibitem[Abdul-Jabbar \latin{et~al.}(2018)Abdul-Jabbar, Poerschke, Gabbett, and
  Levi]{2018_TAbdulJabbar}
Abdul-Jabbar,~N.~M.; Poerschke,~D.~L.; Gabbett,~C.; Levi,~C.~G. Phase
  equilibria in the zirconia–yttria/gadolinia–silica systems. \emph{J. Eur.
  Ceram. Soc.} \textbf{2018}, \emph{38}, 3286--3296\relax
\mciteBstWouldAddEndPuncttrue
\mciteSetBstMidEndSepPunct{\mcitedefaultmidpunct}
{\mcitedefaultendpunct}{\mcitedefaultseppunct}\relax
\EndOfBibitem
\end{mcitethebibliography}

\end{document}